\documentclass[10pt,aps,prd,
tightenlines,twocolumn,
superscriptaddress,
nofootinbib]{revtex4}
\usepackage{hyperref}

\usepackage{amsmath,amsfonts,epsfig,graphicx}
\usepackage{multirow}
\usepackage[latin1]{inputenc}
 \usepackage{amssymb}
 \usepackage{verbatim}
\usepackage{float}
\usepackage{slashed} 
\usepackage{subcaption}
\begin{document}
\title{Loop-induced $ZZ$ production at the LHC: an improved description by matrix-element matching}

\author{Congqiao Li}
\email{congqiao.li@cern.ch}
\affiliation{
Department of Physics and State Key Laboratory of Nuclear Physics and Technology, Peking University, 100871 Beijing, China}

\author{Ying An}
\affiliation{
Department of Physics and State Key Laboratory of Nuclear Physics and Technology, Peking University, 100871 Beijing, China}

\author{Claude Charlot}
\affiliation{Laboratoire Leprince-Ringuet, \'Ecole Polytechnique and IN2P3-CNRS, 91128 Palaiseau Cedex, France}

\author{Roberto Covarelli}
\affiliation{Universit{\`a} di Torino, 10124 Torino, Italy}
\affiliation{INFN Sezione di Torino, 10124 Torino, Italy}

\author{Zhe Guan}
\affiliation{
Department of Physics and State Key Laboratory of Nuclear Physics and Technology, Peking University, 100871 Beijing, China}

\author{Qiang Li}
\affiliation{
Department of Physics and State Key Laboratory of Nuclear Physics and Technology, Peking University, 100871 Beijing, China}

\begin{abstract}
Loop-induced $ZZ$ production can be enhanced by the large gluon flux at the LHC, and thus should be taken into account in relevant experimental analyses. We present for the first time the results of a fully exclusive simulation based on the matrix elements for loop-induced $ZZ + 0,1,2$-parton processes at leading order, matched to parton showers. The new description is studied and validated by comparing it with well-established simulation with jets from parton showers. We find that the matched simulation provides a state-of-the-art description of the final state jets. We also briefly discuss the physics impact on vector boson scattering (VBS) measurements at the LHC, where event yields are found to be smaller by about 40\% in a VBS $ZZjj$ baseline search region, compared to previous simulations. We hence advocate relevant analyses to employ a more accurate jet description for the modeling of the loop-induced process.
\end{abstract}


\maketitle
\section{Introduction}
Pair production of $Z$ bosons is an important background for Higgs boson production or new physics searches at the CERN LHC. The loop-induced gluon-fusion process $gg \to ZZ$~\cite{Binoth:2008pr} contributes formally only at the next-to-next-to-leading order in perturbative quantum chromodynamics (QCD). Nevertheless, it can get enhanced by the large gluon flux at the LHC, and thus should be taken into account in relevant experimental analyses, including, for example, Higgs-boson related measurements using the $ZZ$ decay channel, both in the on-shell~\cite{Sirunyan:2017exp,Aaboud:2017oem} and off-shell regions~\cite{Aaboud:2018puo,Sirunyan:2019twz}, tests of the standard model through diboson inclusive production~\cite{Aaboud:2017rwm,Sirunyan:2018vkx} and vector-boson scattering (VBS)~\cite{Sirunyan:2020alo,Aad:2020zbq}, as well as searches for new physics in various forms of heavy resonances~\cite{Khachatryan:2017wny,Aaboud:2017rel}.

The next-to-leading order (NLO) QCD calculation to loop-induced $ZZ$ production has been evaluated in the gluon--gluon $(gg)$ initial state~\cite{Caola:2015psa, Alioli:2016xab, Caola:2016trd} and in the combined $gg$ and (anti)quark--gluon $(qg)$ initial states~\cite{Grazzini:2018owa}, which overall can increase the Born-level result by an amount ranging from 50\% to 100\%, depending on the renormalization and factorization scale choices. The loop-induced $gg \to ZZg$ cross-section was evaluated in \cite{Campanario:2012bh} and can contribute by more than $10\%$ to the next-to-leading order QCD cross section, in a phase space defined by a jet $p_\mathrm{T}$ threshold of 50 GeV.

On the other hand, experimental analyses employ the full kinematical properties of the events, thus focusing not just on the $ZZ$-related quantities but also on additional jets, both in terms of production rate in fiducial regions and of their phase-space variables. It is therefore crucial to get as precise predictions as possible for such exclusive observables, which include the dijet invariant mass $m_{jj}$ and the absolute dijet pseudorapidity separation $|\Delta\eta_{jj}|$. 
It is possible to get such a full exclusive control
at hadron level on the complex event topology at the LHC, while still reaching approximately next-to-leading logarithmic accuracy, with the help of recent sophisticated matching methods between matrix elements and parton showers~\cite{Catani:2001cc,Alwall:2007fs}.
In Ref.~\cite{Cascioli:2013gfa} a similar loop-induced process $gg \to WW$ associated with 0 and 1 jet is fully simulated at the leading order (LO) and matched to parton showers using \textsc{Sherpa}, and the impact on jet observables compared to the inclusive simulation is discussed. The $0,1,2$-jet simulation and its potential impact on dijet related observables is yet to be explored.

We present here the results of a fully exclusive simulation of gluon-mediated $Z$ pair
production based on the matrix elements for loop-induced $ZZ + 0,1,2$ parton(s) at LO matched to parton showers, where $gg$, $qg$ and (anti)quark--(anti)quark $(qq)$ partonic initial states are all included. We examine and validate this new description by comparing it with established simulations where jets are described from parton showers. We find that the matched simulation provide a state-of-the-art accurate description of the final state jets, and is not in agreement with previous simulation for the jet kinematics. We finally focus on the phase space region with two on-shell $Z$ bosons and discuss the impact on the VBS $ZZ$ measurement. We find a large event yield discrepancy (up to 43\%) using the matched $ZZ$ simulation.

The paper is organized as follows. We begin by describing our methodology, describing the steps of the matrix element generation in Sec.~\ref{sec:me} and the matching to parton showers in Sec.~\ref{sec:ps}. Then we provide the computational details in Sec.~\ref{sec:comp}, and the validation of our results in Sec.~\ref{sec:val}. We show that the matching procedure provides reliable results at the LHC and that the effects are significant. Finally, we discuss briefly the physics impact in Sec.~\ref{sec:phys} and conclude in the last section.

\section{Matrix-Element Simulation} \label{sec:me}

\begin{figure}[!ht]
\begin{center}
\begin{subfigure}{0.20\textwidth} \captionsetup{justification=centering}
\includegraphics[height=0.8\textwidth]{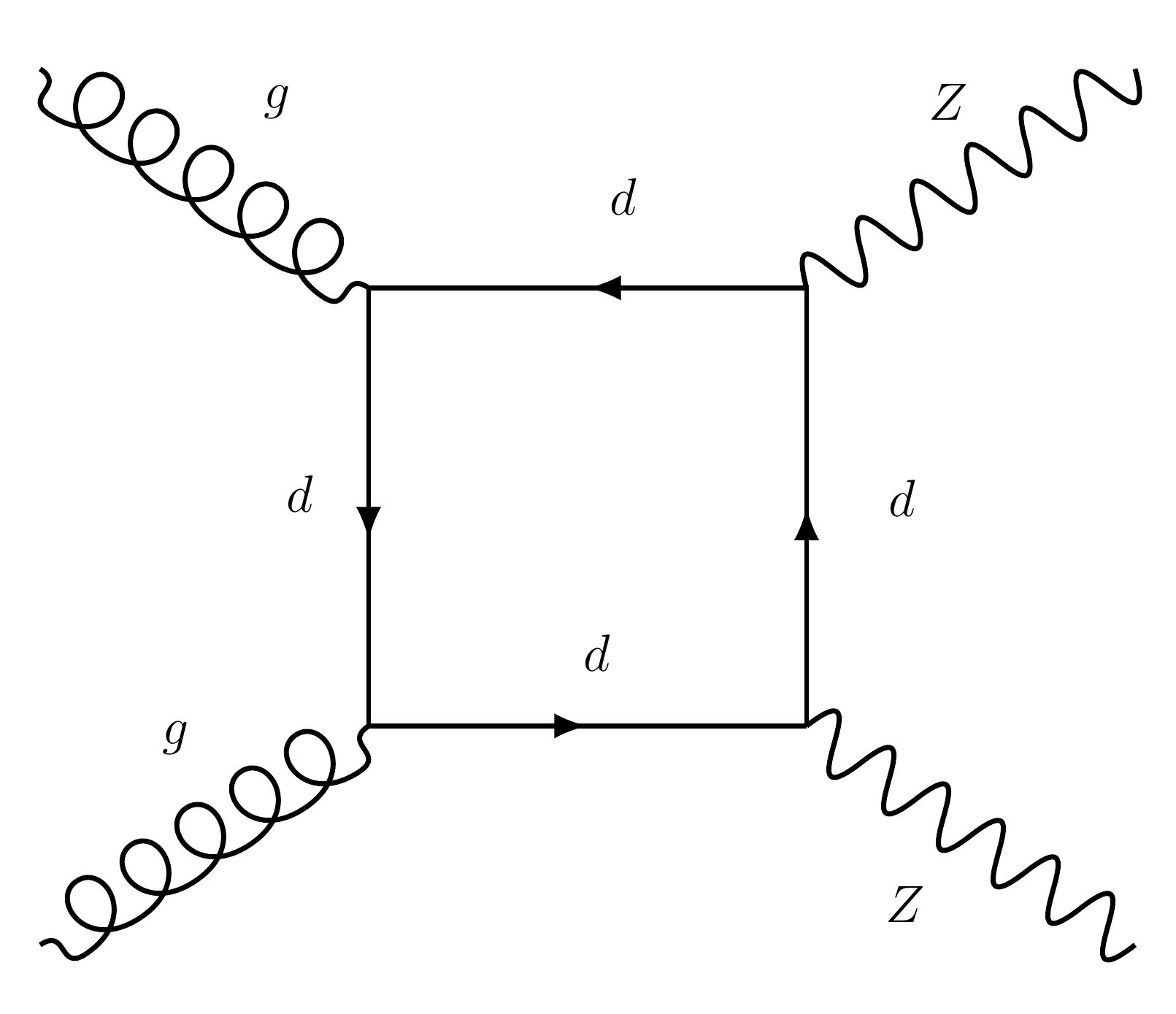} \caption{}
\end{subfigure}
\begin{subfigure}{0.20\textwidth} \captionsetup{justification=centering}
\includegraphics[height=0.8\textwidth]{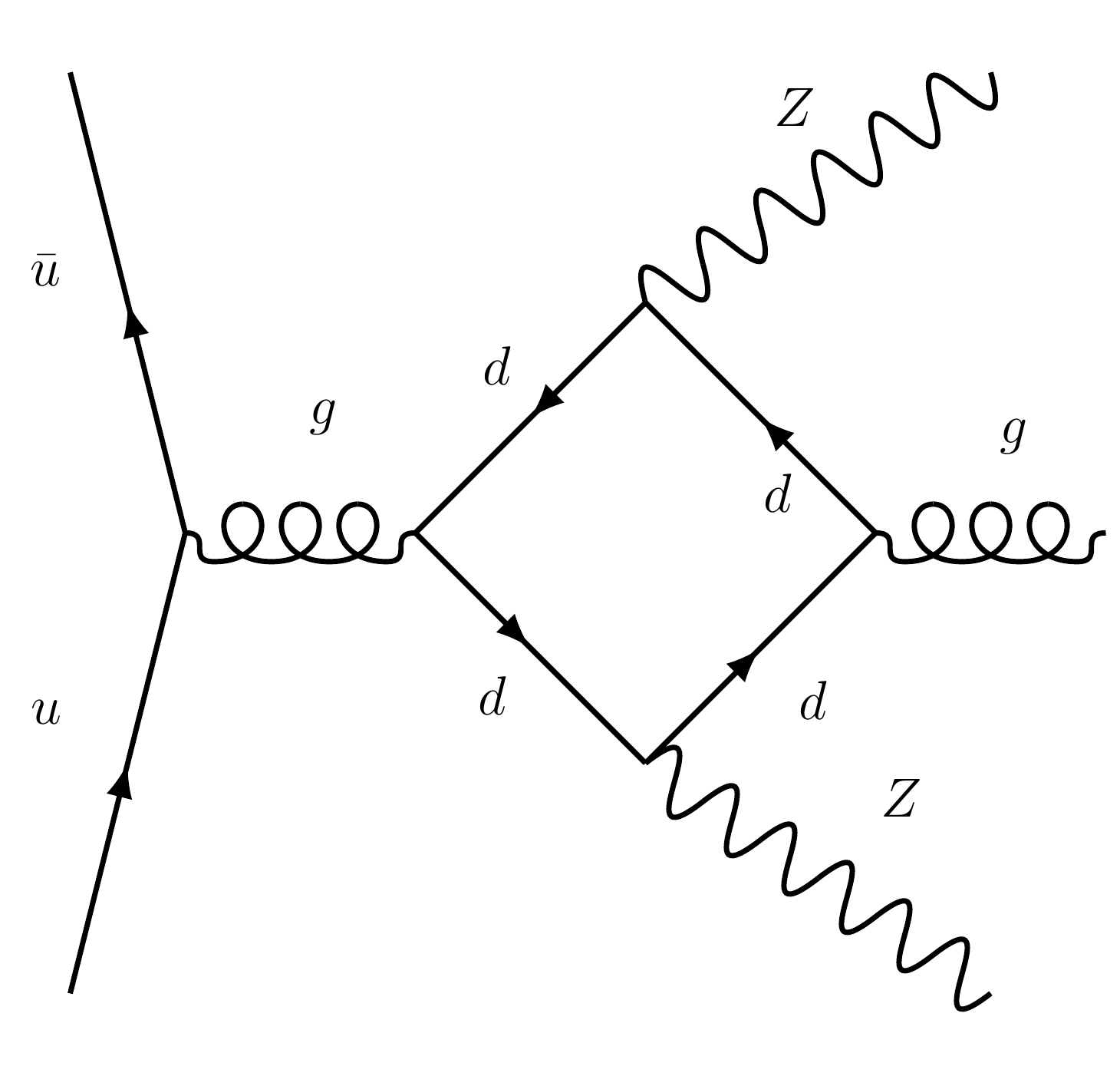} \caption{}
\end{subfigure}\\
\begin{subfigure}{0.20\textwidth} \captionsetup{justification=centering}
\includegraphics[height=0.8\textwidth]{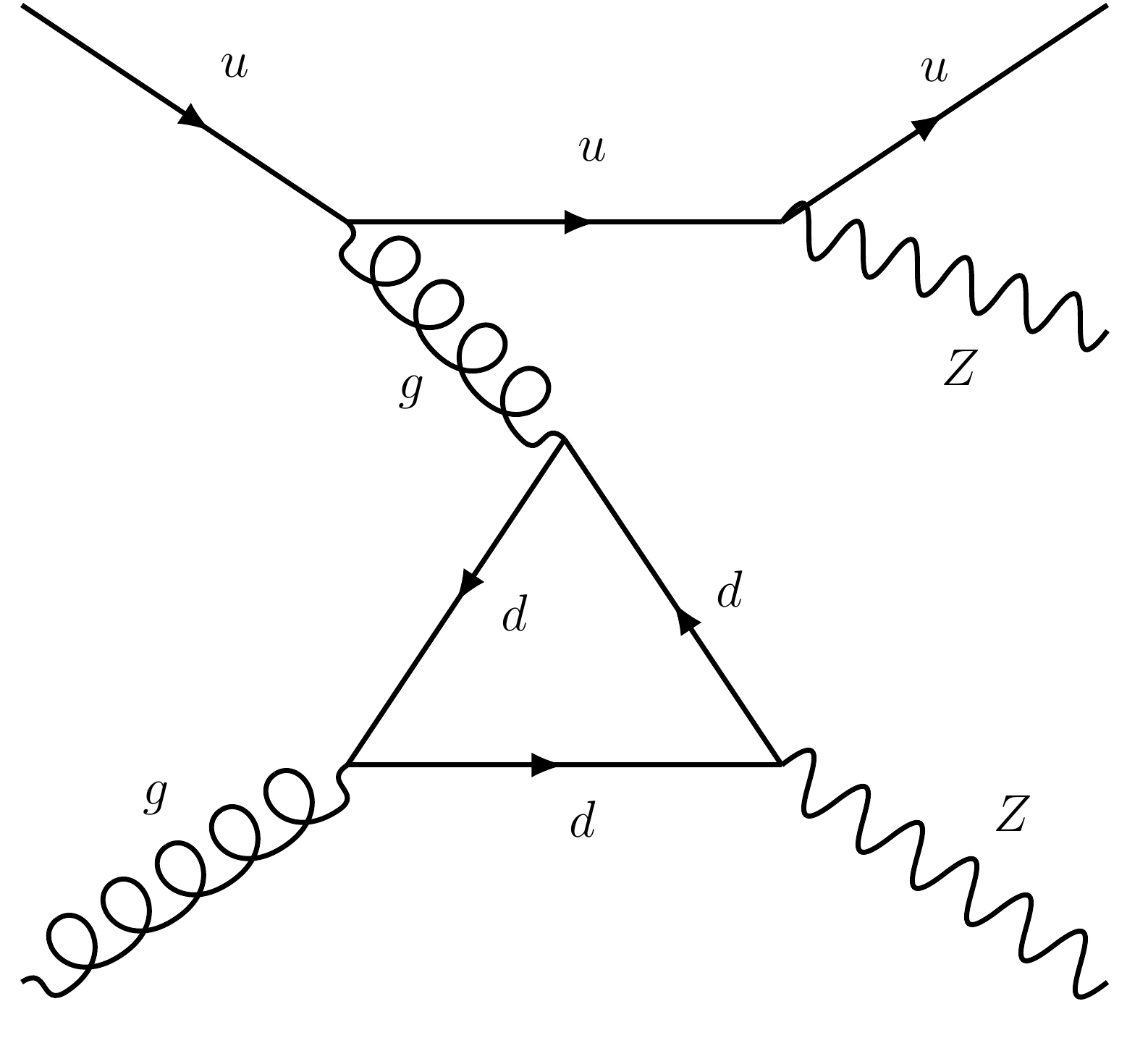} \caption{}
\end{subfigure}
\begin{subfigure}{0.20\textwidth} \captionsetup{justification=centering}
\includegraphics[height=0.8\textwidth]{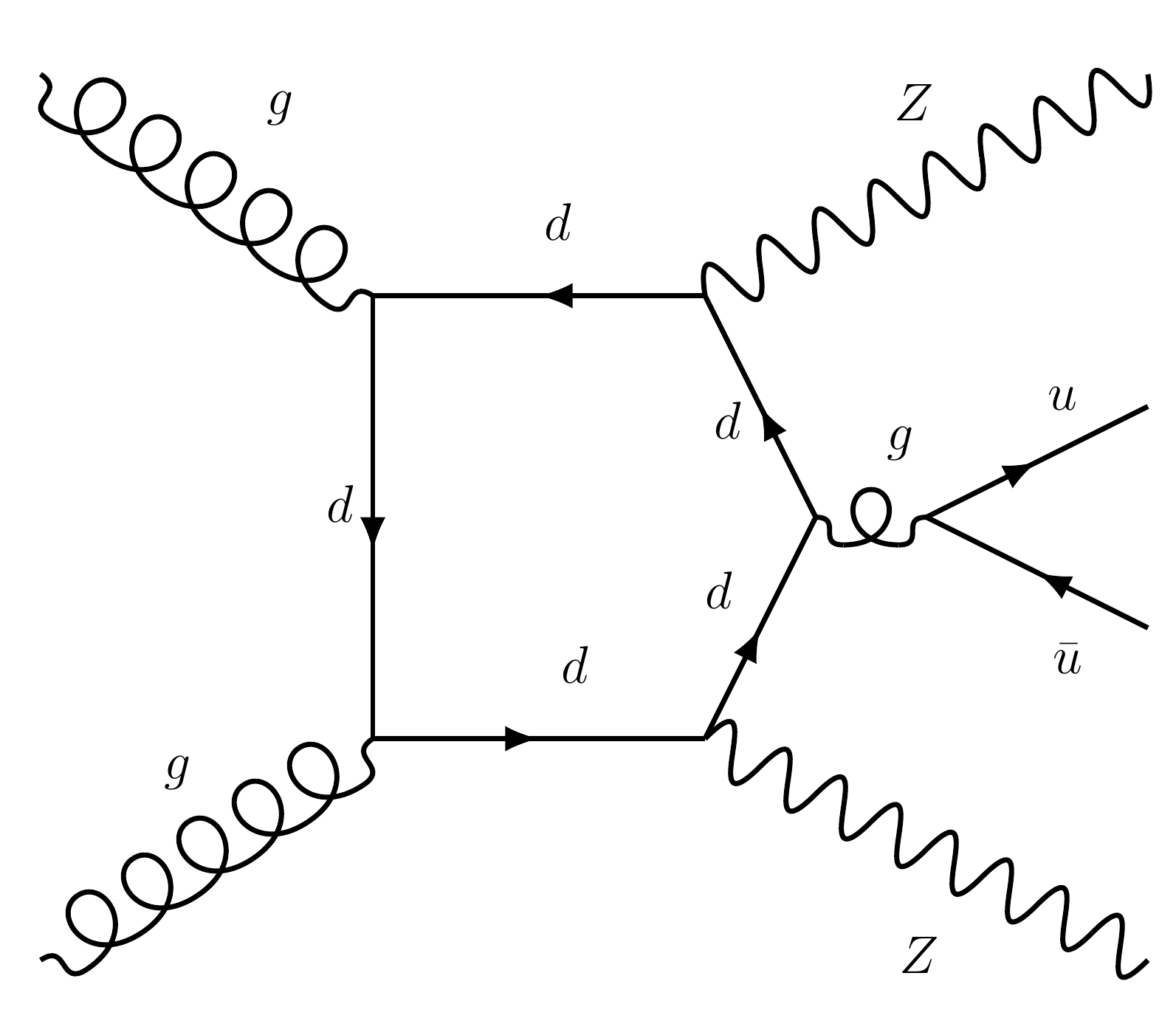} \caption{}
\end{subfigure}\\
\begin{subfigure}{0.20\textwidth} \captionsetup{justification=centering}
\includegraphics[height=0.8\textwidth]{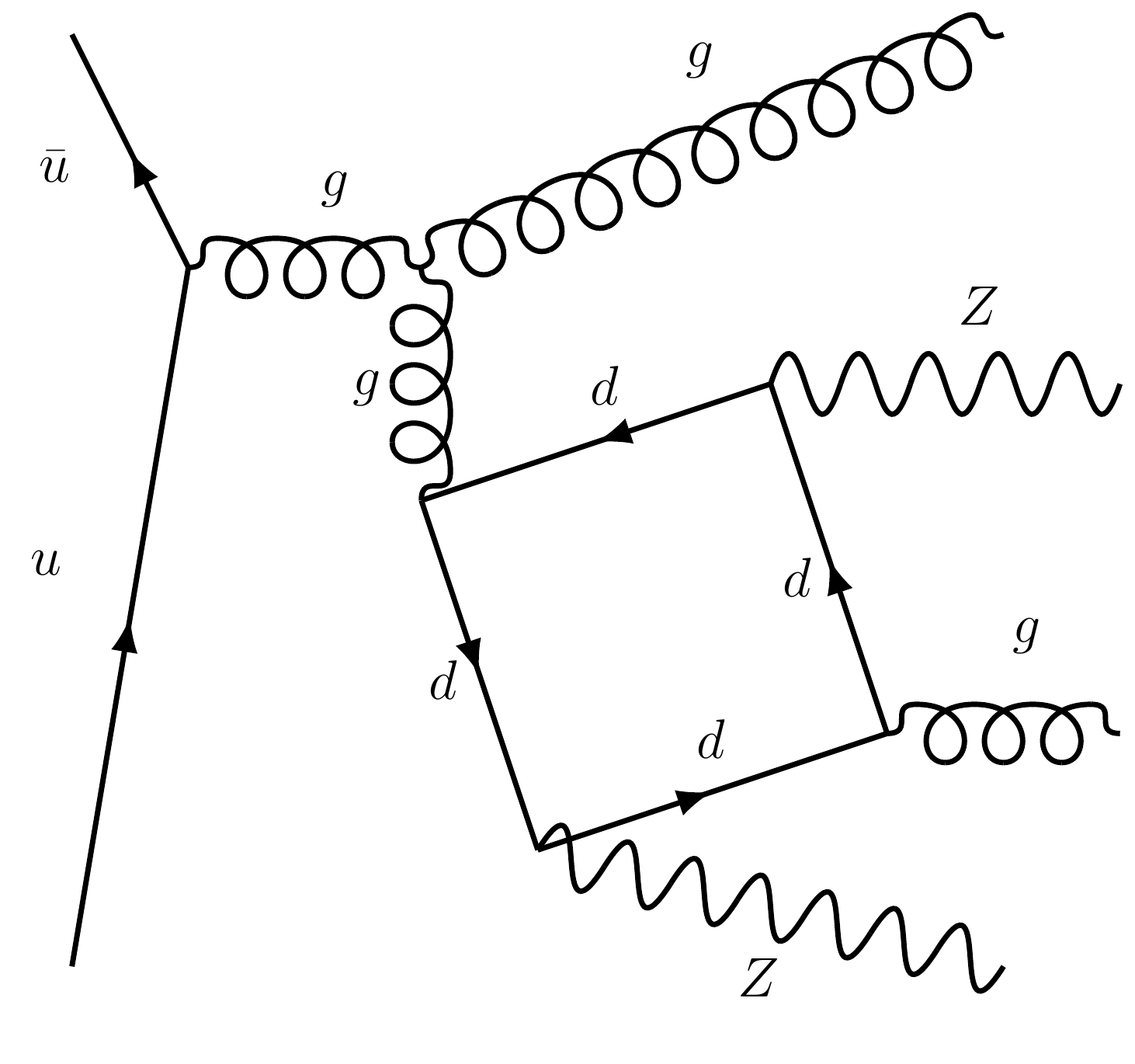} \caption{}
\end{subfigure}
\begin{subfigure}{0.20\textwidth} \captionsetup{justification=centering}
\includegraphics[height=0.8\textwidth]{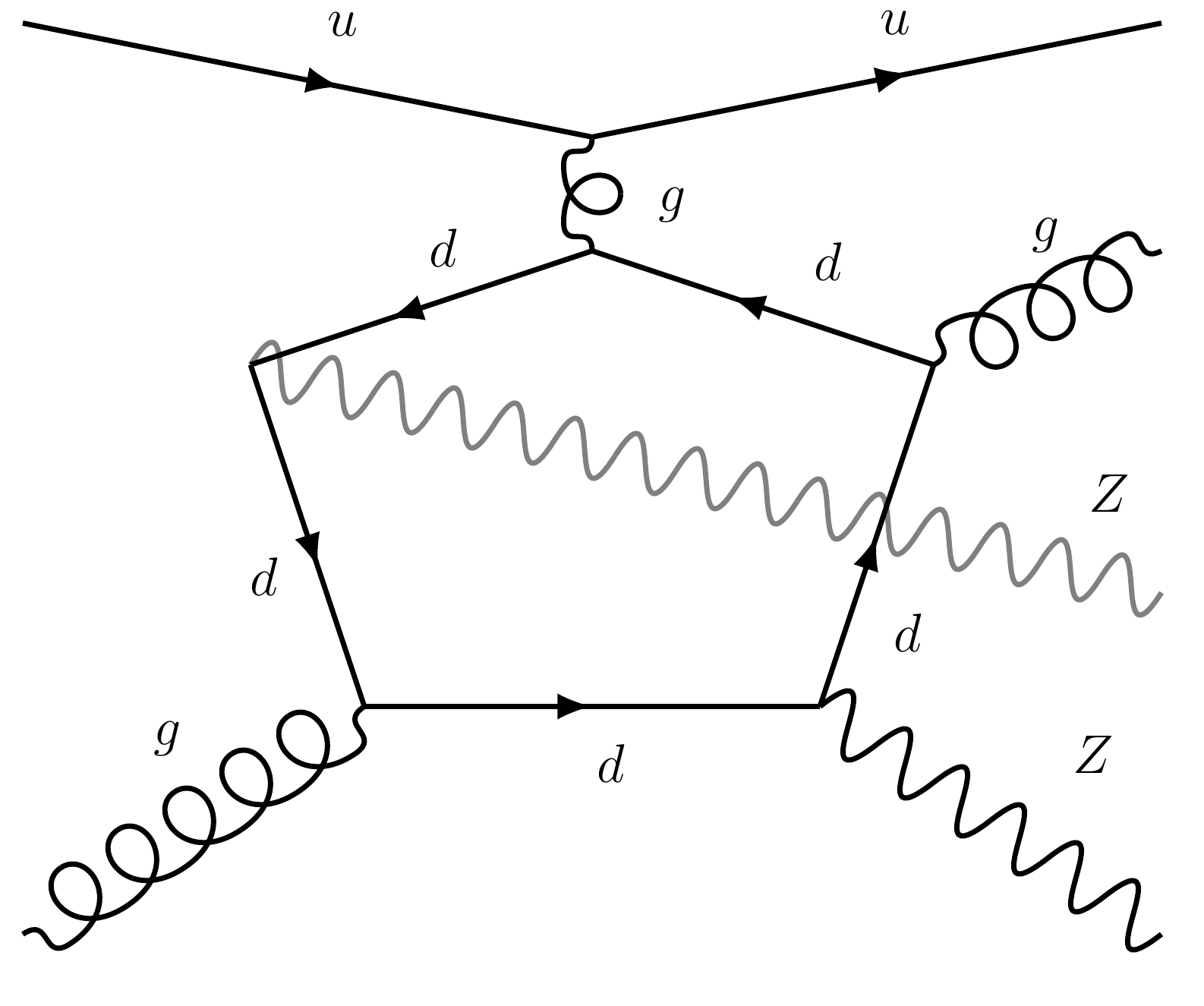} \caption{}
\end{subfigure}
\end{center}
\caption{Example Feynman diagrams for the loop-induced $gg\to ZZ$ process with 0 extra partons (a), 1 parton (b,~c), and 2 partons (d,~e,~f). The $qg$ and $q{\bar q}$ partonic channels are also considered to include the contribution from the initial-state radiation. As pointed out in Sec.~\ref{sec:val}, some 1- and 2-parton sub-processes involve jets that can be emitted directly from the loop (b,~d,~e,~f), thus not fulfilling the definition of initial- or final-state radiation.}
\label{fig:diagram}
\end{figure}

We study $ZZ$ production from the gluon-gluon fusion process which, at the lowest order, is:
\begin{equation}
    gg \rightarrow ZZ + 0,1,2\;\text{jet}
\end{equation}
with the $ZZ$ pair decaying to $\ell^+\ell^-\ell'^+\ell'^-$. The LO for this process contains a quark loop, hence the amplitude for the 0-jet process is at the order of $\mathcal{O}(\alpha_s^2\alpha^4)$ and higher order in $\alpha_s$ for the 1- or 2-jet process. We also take into account the effect of the initial-state radiation (ISR), therefore the $qg$, $qq$ and $gg$ partonic channels are also included in our simulation.
Fig.~\ref{fig:diagram} provides some examples of Feynman diagrams for the 0-, 1- and 2-parton sub-processes, where the $Z$ decay products are omitted.

The loop-induced $ZZ + 0,1,2$ parton processes is simulated with LHC settings, using \textsc{MadGraph5\_aMC@NLO}~\cite{Alwall:2014hca} version 2.6.5. The event sample is generated at LO via a specialized loop-induced mode~\cite{Hirschi:2015iia}, using the commands below.
\begin{verbatim} 
generate g g > z z [noborn=QCD]
add process p p > z z j [noborn=QCD]
add process p p > z z j j [noborn=QCD]
\end{verbatim}
It is known that genuine loop-induced diagrams cannot be automatically selected out of all one-loop diagrams in the \textsc{MadGraph} loop-induced mode. In our case, one-loop diagrams also consist of those serving as one-loop corrections to the tree diagrams, which should be excluded in this simulation, as they do not pertain to the $gg$ initial state. A ``diagram filter'' is especially designed following a suggestion from \textsc{MadGraph} authors~\cite{MLMDISCUSS} to discard those diagrams, using the criteria below.
\begin{itemize}
    \item The loop in the diagram should not contain any gluon line, so that all vertex- and box-correction diagrams are discarded.
    \item The loop in the diagram should be connected to at least one $Z$, $W$ boson or photon, to avoid diagrams representing gluon self-energy corrections through quark lines, and diagrams mediated by a Higgs boson.
\end{itemize}

Once the filter is applied, only genuine $gg \to ZZ$ loop-induced diagrams remain. The Higgs-mediated $gg\to H \to ZZ$ process is also excluded due to the second condition\footnote{The main reason is just computing time saving. Although it is known that the Higgs-mediated $ZZ$ production and its interference with the main $ZZ$ production process plays an important role on the total cross section, we assessed that it has negligible effect on the jet kinematics which is the main topic of this paper. Besides, the benchmark study in Sec.~\ref{sec:phys} requires $Z$ bosons to be on-shell, further diminishing the Higgs contribution.}. For the generation commands, it is worth noting that we specify ``$pp$'' instead of ``$gg$'' as the initial state to include initial-state radiation (ISR), where an initial-state quark can transform to a gluon through radiation, which then takes part in the hard process. The use of ``$pp$'' initial state brings significantly more Feynman diagrams. Only for the 0-jet process, using ``$pp$'' is equivalent to ``$gg$'' since it does not introduce extra loop-induced diagrams. 

The decay of the $ZZ$ pairs to $\ell^+\ell^-\ell'^+\ell'^-$ ($\ell,\,\ell^{\prime}=e,\,\mu$) is implemented in \textsc{pythia} (version 8.230), since the more precise matrix-element based simulator \textsc{MadSpin} is incompatible with the loop-induced mode of \textsc{MadGraph} and a direct generation of the 4-charged lepton final state, including all $Z/\gamma^*$ contributions, features a too complex phase-space integration, hence exceeding current computational possibilities. Therefore, the matrix-element simulation only handles the $ZZ$ plus jets production, leaving both $Z$ at their pole mass, and the spin correlations of the outcoming leptons are thus not simulated.

In the matrix-element simulation, we treat the bottom quark as a massless parton, while taking into account the loop contribution from the massive top quark and the CKM quark-mixing effect. We adopt the on-shell top quark mass $m_t = 173.0\;\mathrm{GeV}$, and the width $\Gamma_t = 1.4915\;\mathrm{GeV}$. The simulation uses the \texttt{NNPDF 3.1} next-to-next-to-leading order parton distribution (PDF) set, with $\alpha_s(m_Z)$ set to 0.118, and adopts the following electroweak parameters:
\begin{equation}
\begin{split}
m_{Z} &= 91.1880\;\mathrm{GeV},\\
\Gamma_{Z} &= 2.441404 \;\mathrm{GeV},\\
G_F &= 1.16639 \cdot 10^{-5} \;\mathrm{GeV}^{-2}.
\end{split}
\quad
\begin{split}
m_{W} &= 80.4190\;\mathrm{GeV},\\
\alpha^{-1} &= 132.5070,\\
&\quad
\end{split} 
\end{equation}

\section{Matching with Parton Showers} \label{sec:ps}
The ``Les Houches'' events (LHE) produced by \textsc{MadGraph} are interfaced to \textsc{pythia} for parton showering. The MLM matching procedure with the shower-$k_{\mathrm{T}}$ scheme~\cite{Alwall:2007fs,Alwall:2008qv} is applied on the $0,1,2$-parton sub-processes to avoid overlapping of the jet phase space as modeled from the matrix elements and the parton showers. The method introduces a cutoff scale $Q^\mathrm{ME}_\mathrm{min}$ at the matrix-element level to remove events with soft partons, and applies another scale $Q^\mathrm{jet}_\mathrm{min}$ in the parton showering step, more specifically onto the $n+1 \to n$ differential jet rates (DJR) of the different $n$-parton sub-process. The $n+1 \to n$ DJR measures the jet splitting rate at a stage when $n$ extra jet(s) are left, during a continuous clustering procedure on the final-state particles. Under such selection defined by $Q^\mathrm{ME}_\mathrm{min}$ and $Q^\mathrm{jet}_\mathrm{min}$, the final event sample is a combined subset of events coming from each $n$-parton sub-process. The jets in each event thus include both harder jets stemming from the matrix-element calculation and softer ones from the parton showers. 

\begin{figure}[!ht]
\begin{center}
\begin{subfigure}{0.5\textwidth} \captionsetup{justification=centering}
\includegraphics[width=0.8\textwidth]{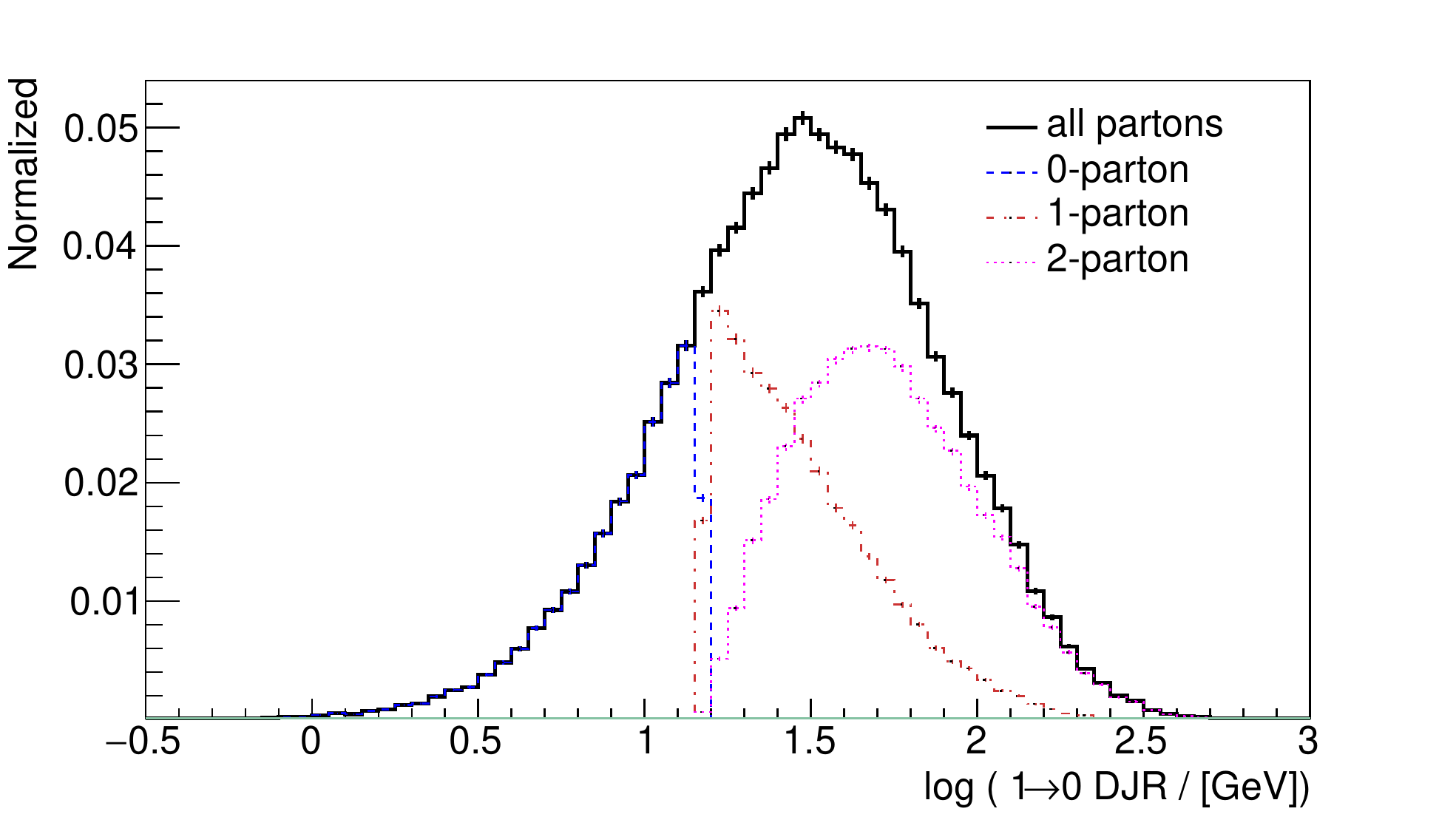}\\
\includegraphics[width=0.8\textwidth]{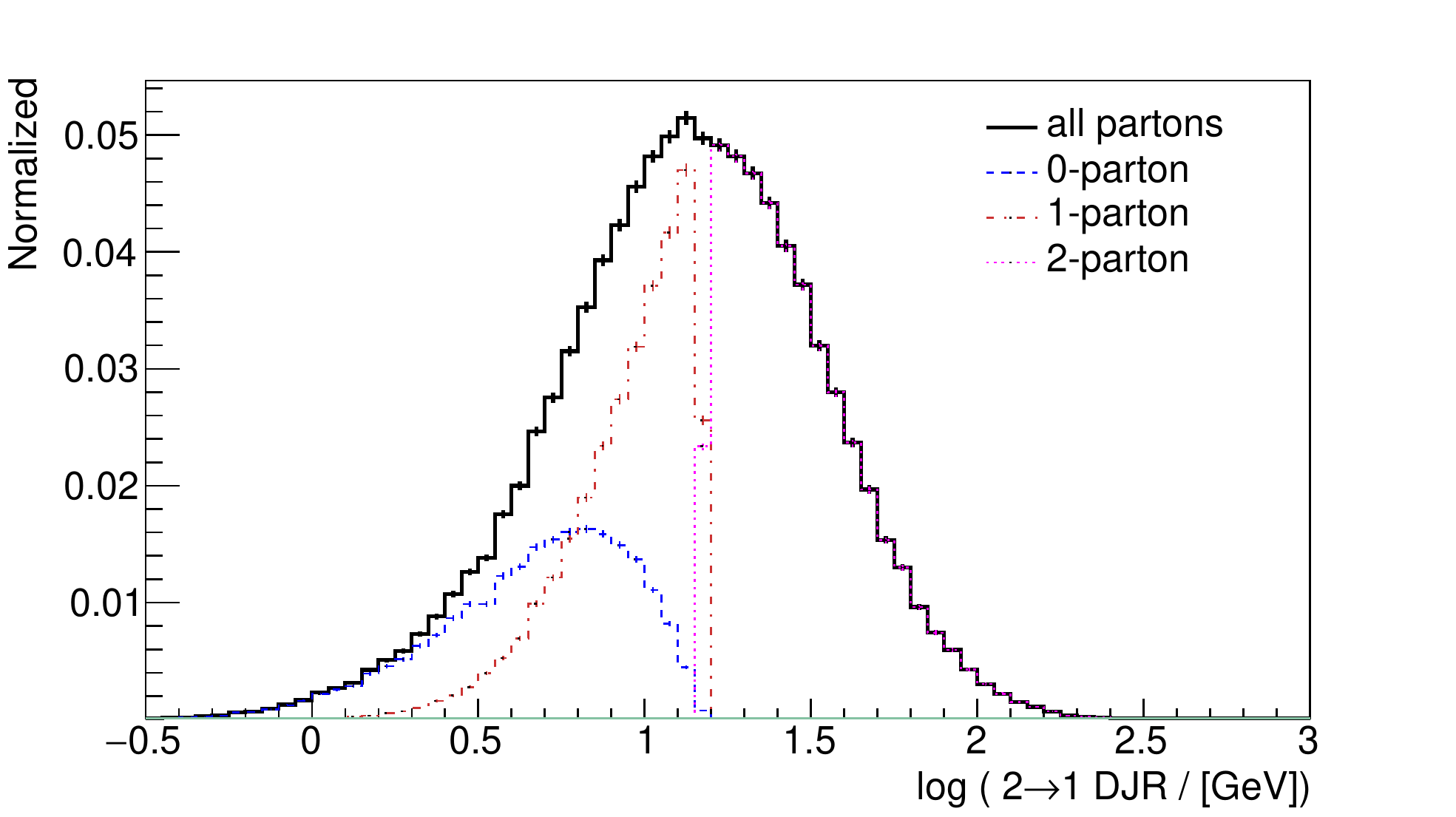}
\caption{}
\end{subfigure} \\
\begin{subfigure}{0.5\textwidth} \captionsetup{justification=centering}
\includegraphics[width=0.8\textwidth]{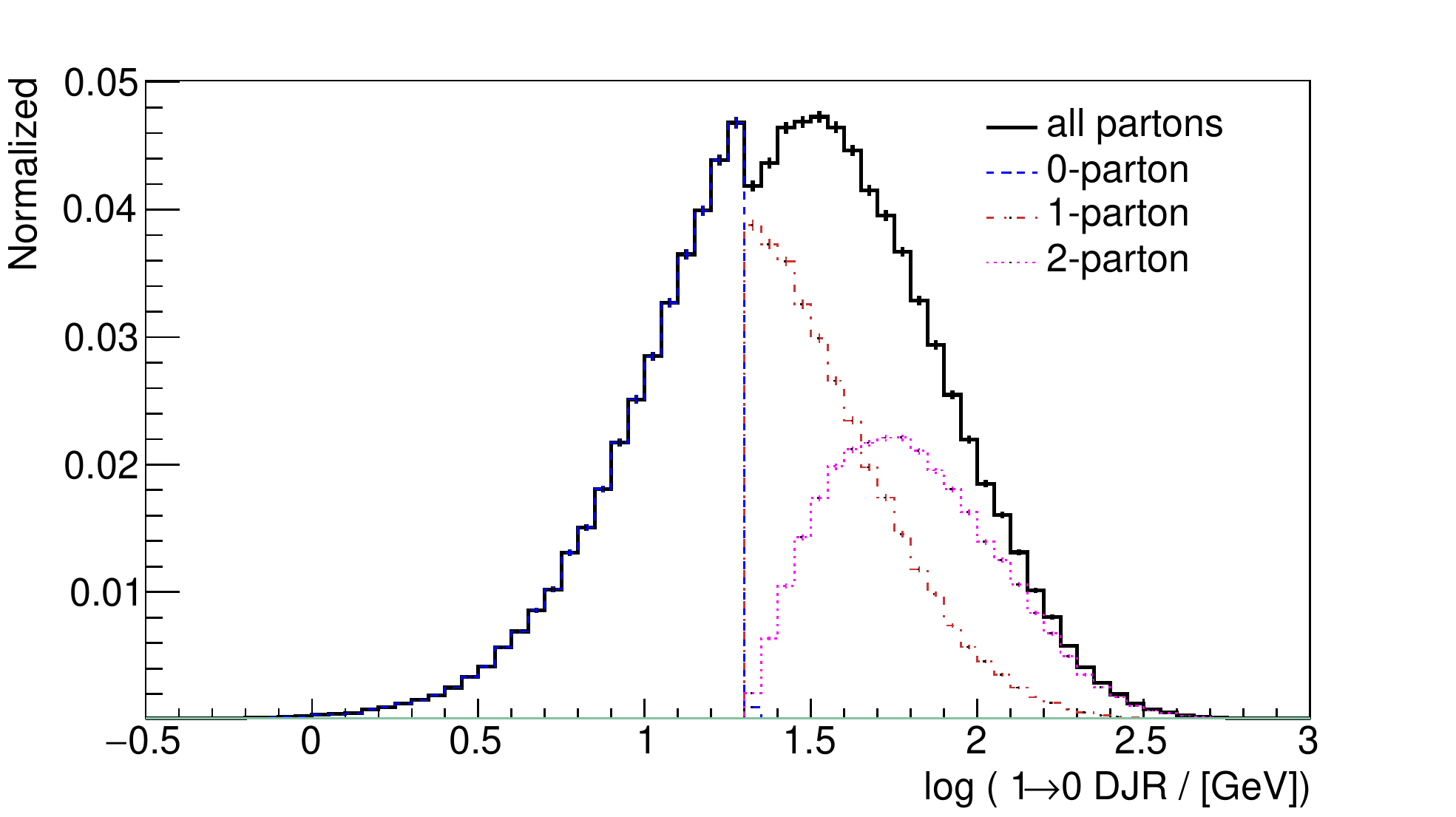}\\
\includegraphics[width=0.8\textwidth]{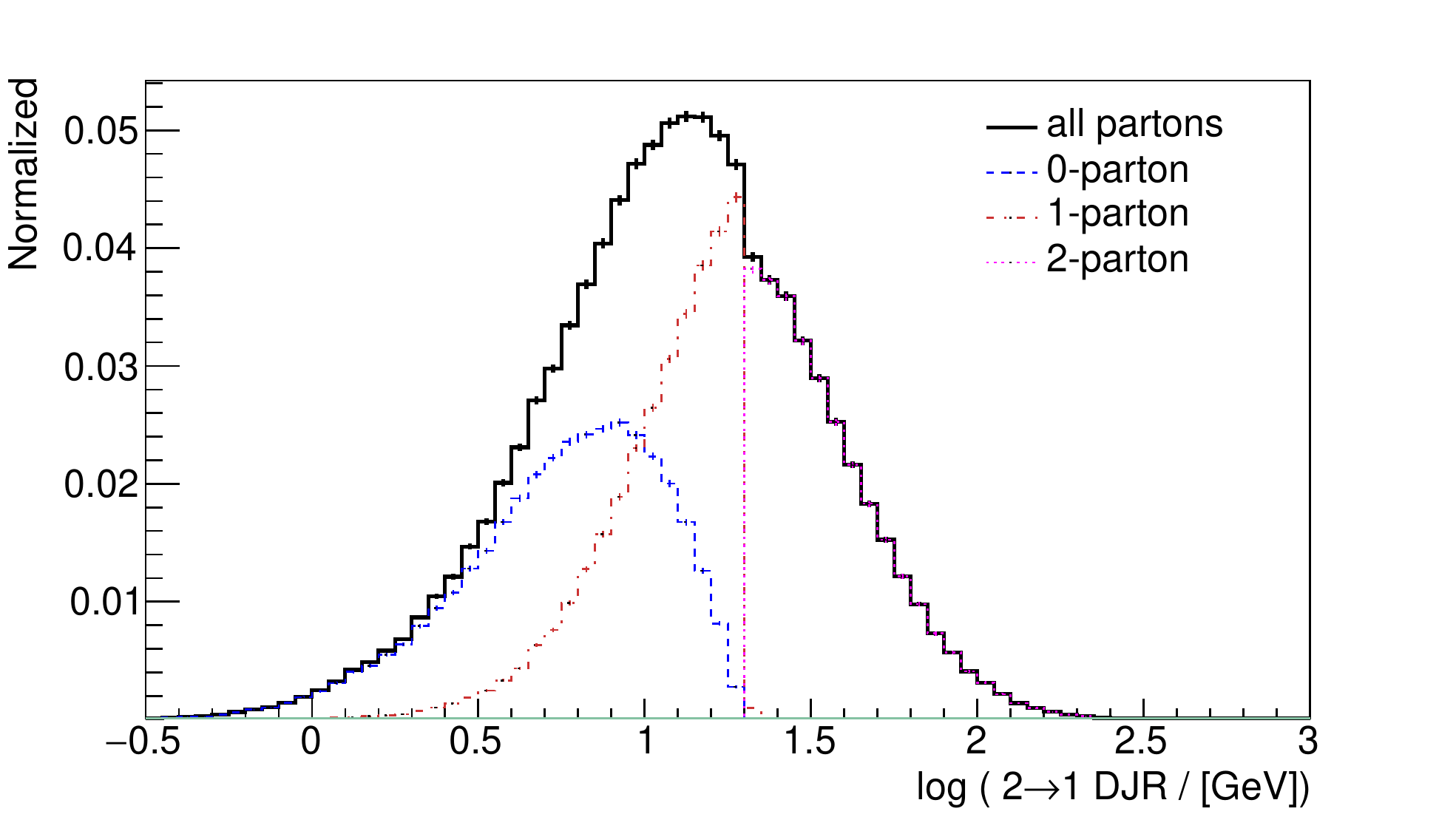}
\caption{}
\end{subfigure}
\end{center}
\caption{The differential jet rate (DJR) distribution for the first and second jet in the loop-induced $gg\to ZZ+0,1,2$ jet(s) process with the MLM cutoff parameters set to $Q^\mathrm{ME}_\mathrm{min}=5$~GeV, $Q^\mathrm{jet}_\mathrm{min}=15$~GeV (a), and another scale choice $Q^\mathrm{ME}_\mathrm{min}=10$~GeV, $Q^\mathrm{jet}_\mathrm{min}=20$~GeV (b) suggested for the $W$ or $Z$ production process.}
\label{fig:djr}
\end{figure}

The optimal $Q^\mathrm{ME}_\mathrm{min}$ value depends on the specific process, while $Q^\mathrm{jet}_\mathrm{min}$ has the default value of $\mathrm{max}\!\left\{Q^\mathrm{ME}_\mathrm{min}+10\,\mathrm{GeV},\,1.2\,Q^\mathrm{ME}_\mathrm{min}\right\}$. 
To validate the matching procedure in the loop-induced $gg\to ZZ+0,1,2$ jet(s) process, the effect of varying the matching cutoff parameters $Q^\mathrm{ME}_\mathrm{min}$ and $Q^\mathrm{jet}_\mathrm{min}$ on several distributions, including the DJR, have been extensively studied. 
We observe that the optimal matrix-element level scale $Q^\mathrm{ME}_\mathrm{min}$ in a loop-induced process is smaller than the suggested choice for the single-$W$ or $Z$ production process in LHC collisions~\cite{MATCHINGSCALE,Alwall:2008qv}. 
Fig.~\ref{fig:djr}~(a) shows the $1 \to 0$ and $2\to 1$ DJR distributions in the logarithmic scale with $Q^\mathrm{ME}_\mathrm{min}$ set to 5~GeV. The smooth transition between curves for the different sub-processes signals a good matching result under such parameters. As a comparison, Fig.~\ref{fig:djr}~(b) shows the less smooth DJR distributions with $Q^\mathrm{ME}_\mathrm{min}$ set to 10~GeV. In each case, $Q^\mathrm{jet}_\mathrm{min}$ is set to the default value, namely 15 and 20~GeV. The later scale choice is validated to provide a smooth stitching in the corresponding tree-level $ZZ+0,1,2$ jet(s) production matched to the \textsc{PYTHIA} parton showers, which is as expected. As a further cross-validation, we investigate the effect of the chosen matching scales, $Q^\mathrm{ME}_\mathrm{min}=5$ and $Q^\mathrm{jet}_\mathrm{min}=15$~GeV, on a similar $gg\to ZZ$ simulation limited to $0$ and $1$ extra partons and verify that matching results are also satisfactory.

We also check that, by choosing the optimal $Q^\mathrm{ME}_\mathrm{min}$ cutoff value rather than the conventional one, the matched cross section for both the $0,1,2$-jet and $0,1$-jet processes agrees better with the inclusive result. Besides, we find that moving from the default \textsc{MadGraph} dynamical scale choice, i.e. $\mu$ equal to the central $m_\mathrm{T}$ after $k_\mathrm{T}$ clustering on LHE-level final states~\cite{SCALECHOICE}, to $\mu=m_{ZZ}$ does not influence the goodness of the matching. This enables the application of an NLO/LO $k$-factor correction, which is evaluated using the latter scale choice~\cite{Caola:2015psa}.

\section{Computing Performance} \label{sec:comp}

The matrix-element level simulation of a loop-induced diboson process with up to two extra partons is implemented for the first time. Because of the complexity of the loop calculation and the large number of diagrams compared to a tree-level process, the event generation turns out to be very time-consuming. We use the \textsc{MadGraph} ``gridpack'' mode to produce this sample, as it separates the phase-space integration and event Monte Carlo simulation into two steps, making the complicated process easier to handle. On a local cluster with 184 CPU cores, the \textsc{MadGraph} gridpack production takes 24 hours to generate the matrix element code for a loop-induced $ZZ + 0,1,2$ parton(s) process (on one core only), which contains 24,066 loop diagrams in total. This is followed by a 84-hour concurrent run for the multi-channel phase-space integration. The total time spent in numerical integration for the $0,1,2$-jet matched process is 15300 core-hours, as a comparison with 10.9 core-hours for the $0,1$-jet process, and 0.085 core-hour for the 0-jet process~\footnote{The simulation is run on 2.4~GHz Intel Xeon E5-2680 CPUs.}. We note that the most computationally expensive channel is one of the twelve sub-processes contributing to the $gg \to ZZgg$ simulation, which takes 650 core-hours to complete.

The events are thereafter generated with the use of the gridpack. Because of the complex phase-space topology, event generation is even more expensive in time: the raw LHE event production rate is 8 min/event, which, when considering a MLM matching rate of about 8\%, reaches a net production rate of 100 min/event.

\section{Validation} \label{sec:val}

The MLM matched simulation has an intrinsic advantage in describing the jet phase space. As can be seen from the Feynman diagrams in Fig.~\ref{fig:diagram} (b,~d,~e,~f), the emitted jets in a loop-induced 1- or 2-parton event do not only consists of ordinary ISR and final-state radiation jets, but also involve emissions directly from the loop---a unique feature of the loop-induced process. Simulation of such emissions is beyond the scope of a parton-shower generator (e.g. \textsc{pythia}), and can only be handled by the matrix-element calculation. Thus, this MLM matched simulation is expected to provide the most state-of-the-art modeling of $gg\to ZZ$ process in the dijet phase space simulation.

In order to validate this new description, we compare it with four additional $gg\to ZZ$ simulations. The first is obtained from the \textsc{mcfm} program by calculating $gg\to ZZ\to \ell^+\ell^-\ell'^+\ell'^-$ ($\ell,\ell^{\prime}=e,\,\mu$) matrix elements~\cite{Boughezal:2016wmq}; the second uses \textsc{MadGraph} for $gg\to ZZ$ inclusive simulation with $Z$ bosons decayed by \textsc{pythia}; the third uses \textsc{MadGraph} to simulate only the $gg\to ZZ+1$ parton final state with the same treatment of decays; and the fourth also produced with \textsc{MadGraph} but featuring a $gg\to ZZ+1$ parton simulation matched to the 0-parton process, using the same cutoff parameters as optimized in Sec.~\ref{sec:ps}\footnote{The same treatment is made for the compared \textsc{MadGraph} simulations with the studied $gg\to ZZ+0,1,2$-jet simulation, including the application of the diagram filter and the use of ``$pp$'' initial state.}. 
The \textsc{MadGraph} samples adopt the same definition of parameters, scales and PDF as described in Sec.~\ref{sec:me}.
All simulations use \textsc{pythia} for parton showering.
Therefore, in the first and second simulation extra jets are modeled fully by a  parton-shower approach via \textsc{pythia}, while the remaining simulations have part of the jets (mostly sub-leading in transverse momentum) modeled by \textsc{pythia}. 

\begin{figure*}[!ht]
\begin{center}
\includegraphics[width=0.38\textwidth]{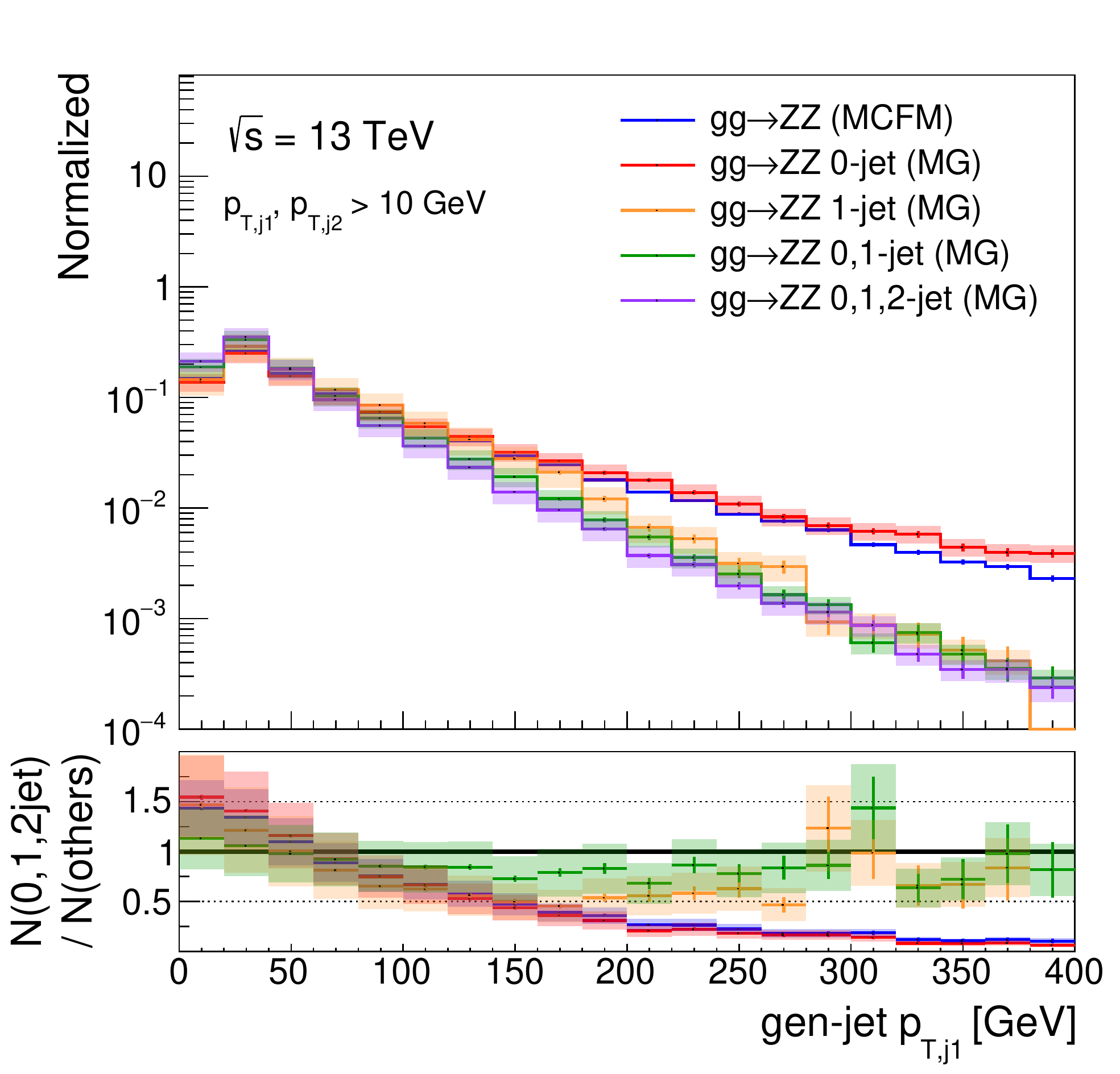}
\includegraphics[width=0.38\textwidth]{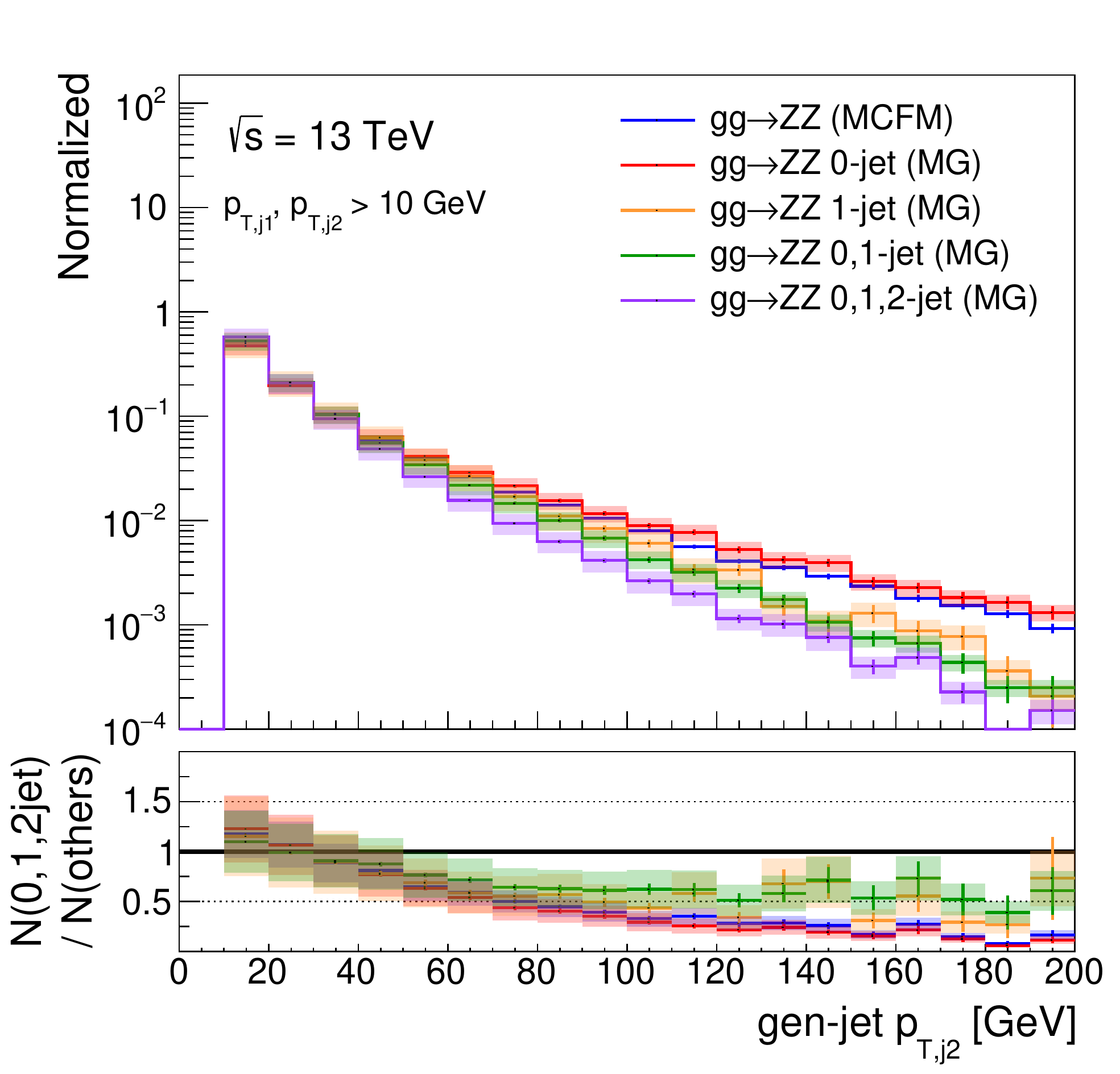}\\
\includegraphics[width=0.38\textwidth]{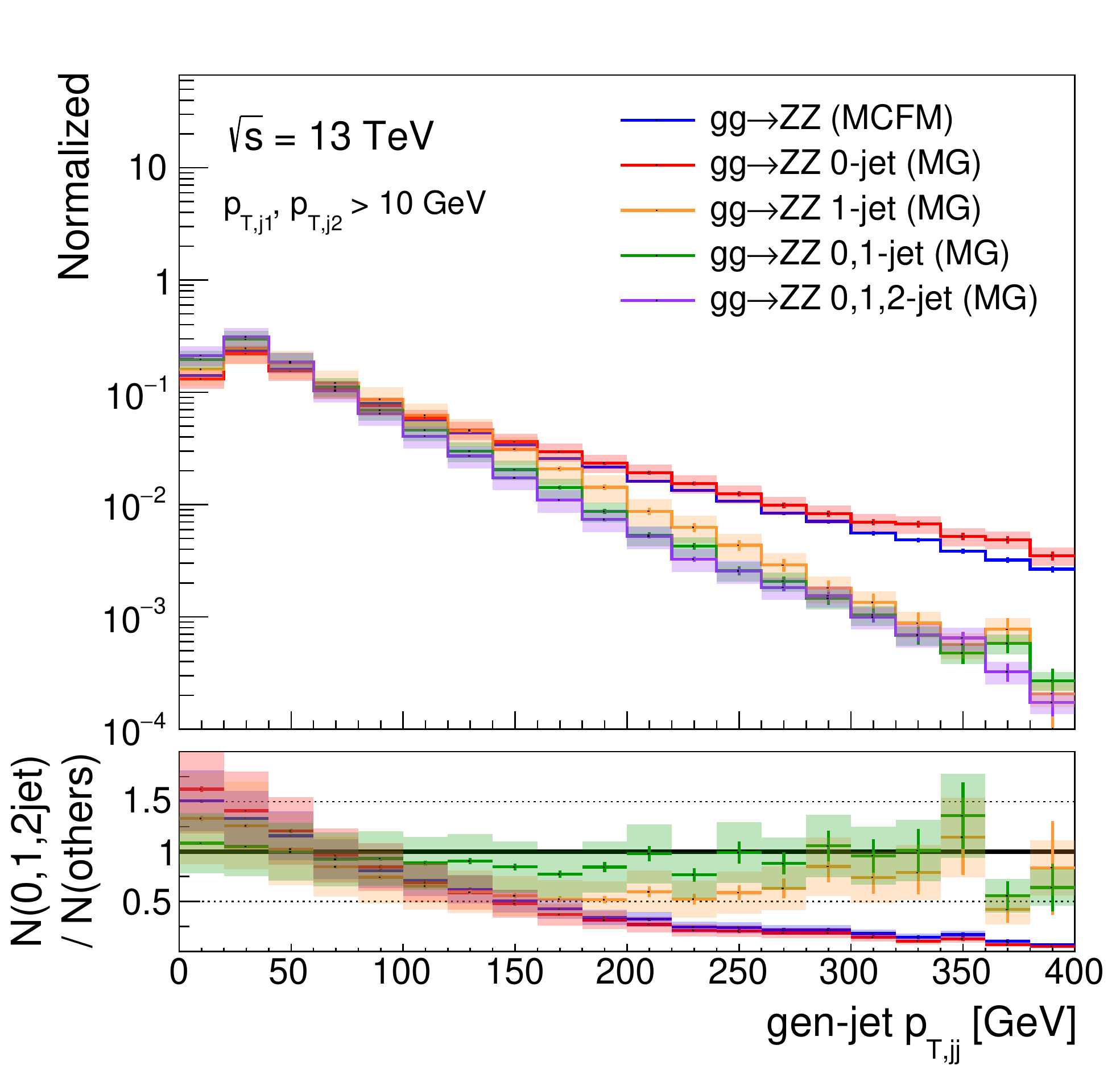}
\includegraphics[width=0.38\textwidth]{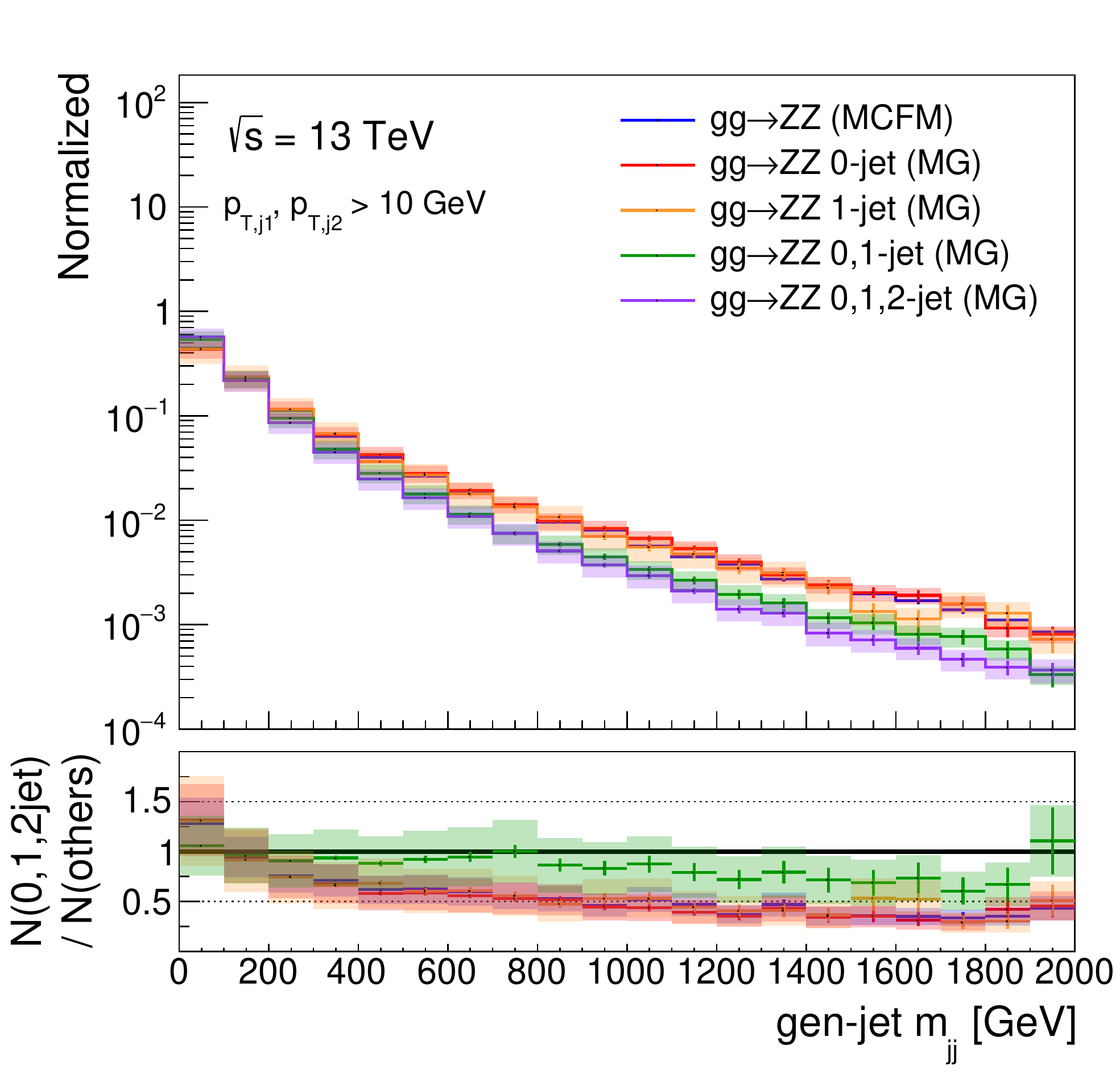}
\end{center}
\caption{The comparison of various generator-level jet observables among the \textsc{mcfm} $gg\to ZZ$ simulation (blue), the \textsc{MadGraph} $gg\to ZZ$ 0-jet simulation (red), 1-jet simulation (orange), $0,1$-jet MLM matched simulation (green), and finally the $0,1,2$-jet matched simulation (purple) as introduced in this work. $p_{\mathrm{T},j_{1,2}} > 10$~GeV is applied as a pre-selection. The error bars indicate the statistical uncertainties. The shaded regions show the combined uncertainties from QCD scales and from the PDF (dominated by the former). The increasing softness of jet kinematics going from the $0$-jet to the $0,1,2$-jet matched simulations can be observed.}
\label{fig:gen_val_plot}
\end{figure*}

We compare those simulations for various generator-level jet observables, where jets are reconstructed from final-state particles, adopting an anti-$k_{\mathrm{T}}$ algorithm with a distance parameter of 0.4. Fig.~\ref{fig:gen_val_plot} shows the spectrum of the generator-level leading and sub-leading jet transverse momenta ($p_{\mathrm{T},j_1}$, $p_{\mathrm{T},j_2}$), the dijet transverse momentum $p_{\mathrm{T},jj}$, and the dijet invariant mass $m_{jj}$. We first notice that the \textsc{MadGraph} and \textsc{mcfm} simulation with pure parton-shower jets agree well in shape. It is of interest to see that, starting from the \textsc{MadGraph} 1-jet simulation, the jet $p_T$ and mass distributions gradually turn softer, which is a consequence brought by the matrix-element modeled jets. Meanwhile, comparing the \textsc{MadGraph} $0,1$-jet matched simulation with the $0,1,2$-jet one, we observe similar behavior in the first jet kinematics, but slight discrepancies in the second jet. This is in agreement with our expectations since the in $0,1$-jet simulation the leading jet is modeled by the matrix-element, similarly to the $0,1,2$-jet simulation. It turns out that the $0,1,2$-jet simulation gives the softest jets, while it should be the most realistic description amongst all the methods. The shaded region in Fig.~\ref{fig:gen_val_plot} represents the combined uncertainties from renormalization and factorization scale (dominant source) and from the parton distribution function, which do not allow to account for the shape differences. The distributions illustrate the sizable discrepancy in dijet phase space modeling between an $0,1,2$-jet or $0,1$-jet MLM matched description and a full parton-shower description.

It is important to note that the majority of the $gg\to ZZ$ loop-induced simulations as implemented in various experimental LHC analyses, have the jets modeled in a full parton-shower approach~\cite{Aaboud:2017oem,Sirunyan:2018vkx,Sirunyan:2017fvv,Khachatryan:2017wny,Aaboud:2017rel} or with some approximation using the similarity of the $H\to ZZ^*$ process~\cite{Sirunyan:2017exp,Sirunyan:2019twz}, although the latter only includes $Z$ off-shell contributions.
Some analyses~\cite{Aaboud:2017rwm,Aaboud:2018puo,Aad:2020zbq} apply an alternative approach, i.e., using \textsc{Sherpa} for the matrix-element simulation of $ZZ$ with zero or one extra jet and match them with the \textsc{Sherpa} parton showers, based on the study in Ref.~\cite{Cascioli:2013gfa}, but this has not yet reached the accuracy of dijet matrix-element simulation. Thus, the discrepancy in Fig.~\ref{fig:gen_val_plot} should be considered carefully in the relevant analyses.

\section{Physics Impact} \label{sec:phys}

The innovation from the earlier, less accurate, modeling of the dijet phase space to the new $0,1,2$-jet matched description may bring potential impacts in relevant analyses. We consider a typical VBS $ZZ$ measurement at the LHC assuming an integrated
luminosity of $150~\mathrm{fb}^{-1}$ as an example and discuss the impact based on generator-level simulation. Referring to the object reconstruction and event selection strategy in the $\ell\ell\ell\ell jj$ channel in the ATLAS and CMS study~\cite{Sirunyan:2020alo,Aad:2020zbq}, we design an algorithm to select four generator-level leptons and determine two lepton-pair candidates, based on their proximity to the $Z$ pole mass. Experimental analyses for $ZZ$ final states require a $Z$-window selection, thus we impose the selection of $60~\mathrm{GeV}<m_{Z_{1,2}}<120~\mathrm{GeV}$ and $m_{ZZ}>160~\mathrm{GeV}$ as the on-shell requirement which makes the on-shell \textsc{MadGraph} simulation still viable. 
On top of that, a $ZZjj$ baseline selection is designed to select the signal, imposing a jet requirement on the leading and sub-leading jets, namely $p_{\mathrm{T},j_{1,2}} > 30$~GeV and $m_{jj} > 100$~GeV. A VBS-enriched region is also defined to increase the VBS signal purity, further requiring $m_{jj} > 400$~GeV and $|\Delta\eta_{jj}|>2.4$.

The $0,1,2$-jet MLM matched simulation is studied at generator level by comparing with the \textsc{mcfm} and \textsc{MadGraph} description with jets from partons showers, with the \textsc{MadGraph} 1-jet simulation, and with the $0,1$-jet matched simulation. 
Table~\ref{tab:yields} shows the yields with the combined scale and PDF uncertainties, and the statistical uncertainties (based on the assumed integrated luminosity) for the four simulated samples after the $ZZjj$ baseline and VBS-enriched selection respectively. We see that the event yields after the $ZZjj$ baseline selection decrease by 43\% for this $0,1,2$-jet description and, less significantly, by a factor of 34\% for the $0,1$-jet description and 9\% for the 1-jet description. The event reduction reaches up to 56\% when moving to the tighter VBS-enriched selection. A $pp\to ZZ+0,1,2$-jet process produced by \textsc{MadGraph} at the LO and at tree level (including no loops) is also presented for event yield comparison. It can be seen that, the proportion of the loop-induced background becomes larger in the VBS-enriched selection, therefore the yield decrease observed in the $0,1,2$-jet matched simulation becomes more relevant in experimental analyses targeting this phase space.
\begin{table}[!ht]
\centering
\caption{Event yields comparison for the \textsc{mcfm} $gg\to ZZ$ simulation, the \textsc{MadGraph} (\textsc{MG}) $gg\to ZZ$ 0-jet simulation, 1-jet simulation, $0,1$-jet MLM matched simulation, the $0,1,2$-jet matched simulation, and the Born-level $pp\to ZZ+0,1,2$-jet matched simulation, after the $ZZjj$ baseline selection and the VBS-enriched selection, respectively. The \textsc{MadGraph} samples for $gg\to ZZ$ with jets are normalized to the \textsc{mcfm} cross section after passing the on-shell requirement of reconstructed $Z$ bosons; the $pp\to ZZ+0,1,2$-jet process is scaled with respect to the $gg\to ZZ+0,1,2$-jet process based on their cross section ratio obtained from \textsc{MadGraph} at LO. All samples are normalized to an integrated luminosity of $150~\mathrm{fb}^{-1}$. The first error term shows the statistical uncertainty (from the assumed data set size), and the second term gives a combined scale and PDF uncertainties as the systematic uncertainty (available for \textsc{MG} samples).}
\begin{tabular}{lcc}
\hline\hline
Process & $ZZjj$ baseline & VBS-enriched \\
\hline
\textsc{mcfm} 0-jet & $98.0 \pm 9.9$  & $26.1 \pm 5.1$  \\
\textsc{MG} 0-jet & $103.1 \pm 10.1 \pm 18.9$  & $27.8 \pm 5.2 \pm 5.1$  \\
\textsc{MG} 1-jet & $88.2 \pm 9.4 \pm 24.5$  & $25.0 \pm 5.0 \pm 6.9$  \\
\textsc{MG} $0,1$-jet & $64.3 \pm 8.0 \pm 12.7$  & $13.5 \pm 3.6 \pm 2.8$  \\
\textsc{MG} $0,1,2$-jet & $55.4 \pm 7.4 \pm 12.5$  & $11.5 \pm 3.3 \pm 2.9$  \\
\hline
\textsc{MG} $pp\to ZZ$ & $586.3 \pm 24.2 \pm 32.1$  & $65.6 \pm 8.1 \pm 5.2$  \\
\hline\hline
\end{tabular}
\label{tab:yields}
\end{table}

\begin{figure*}[!ht]
\begin{center}
\includegraphics[width=0.38\textwidth]{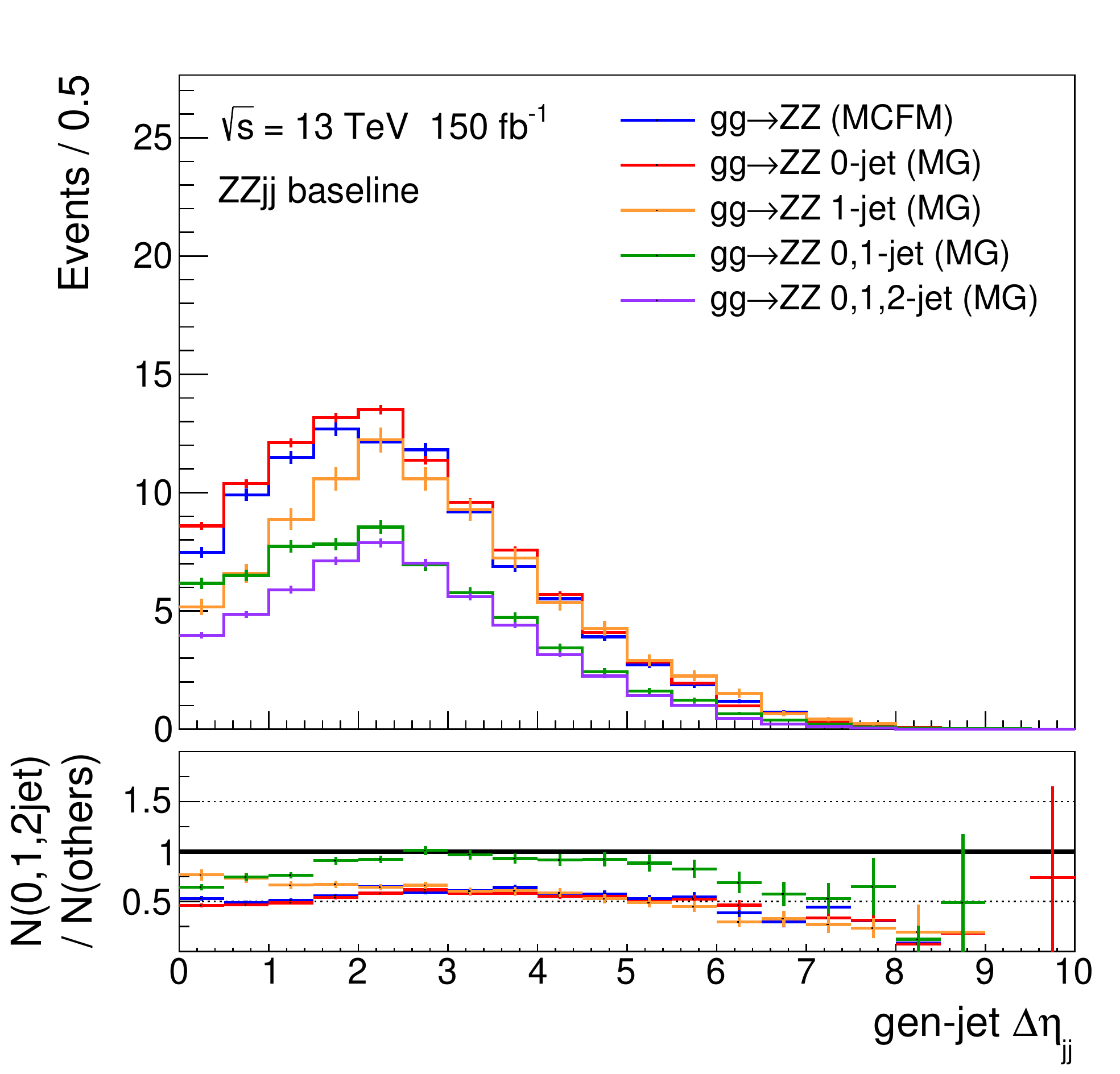}
\includegraphics[width=0.38\textwidth]{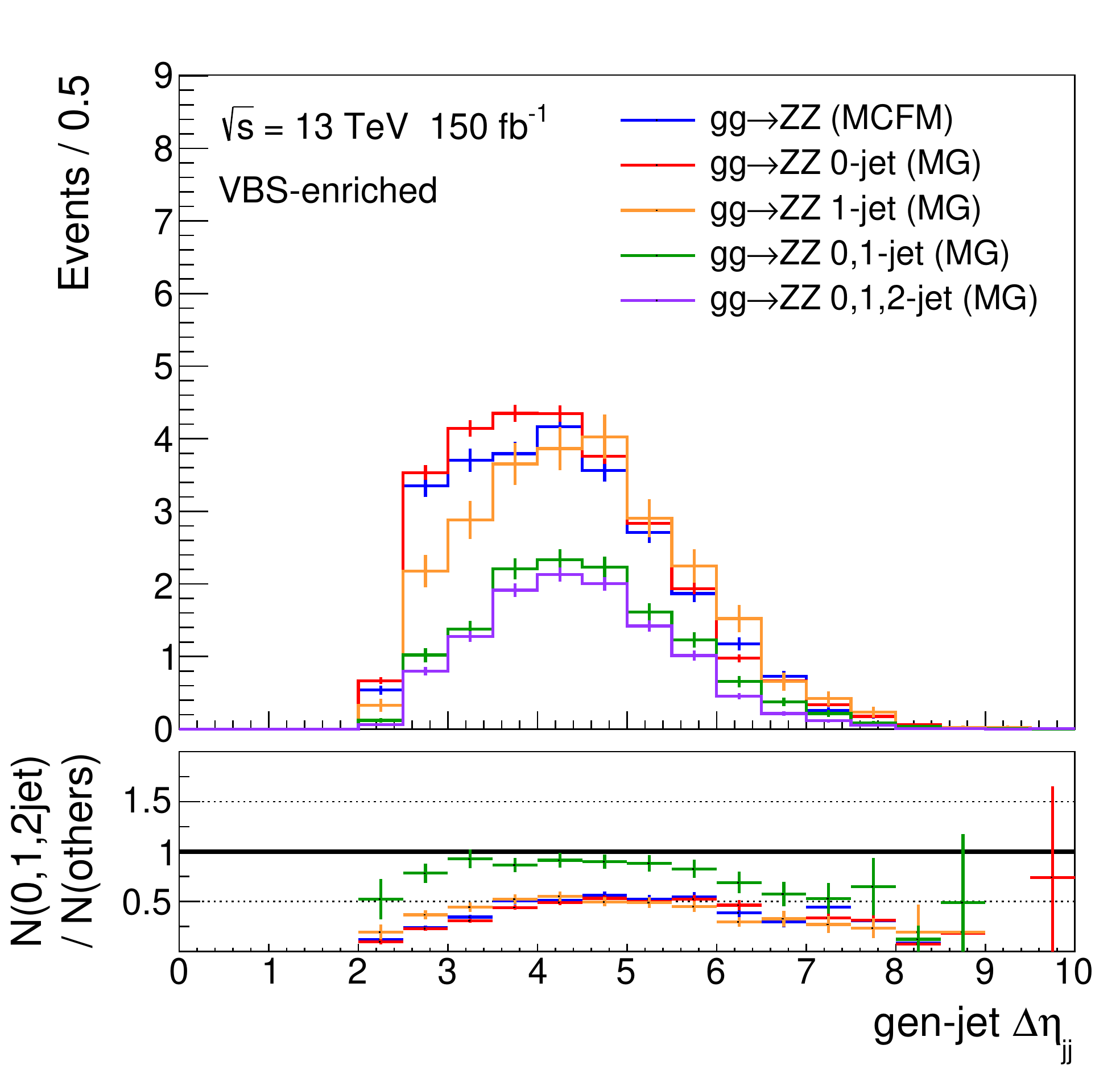}\\
\includegraphics[width=0.38\textwidth]{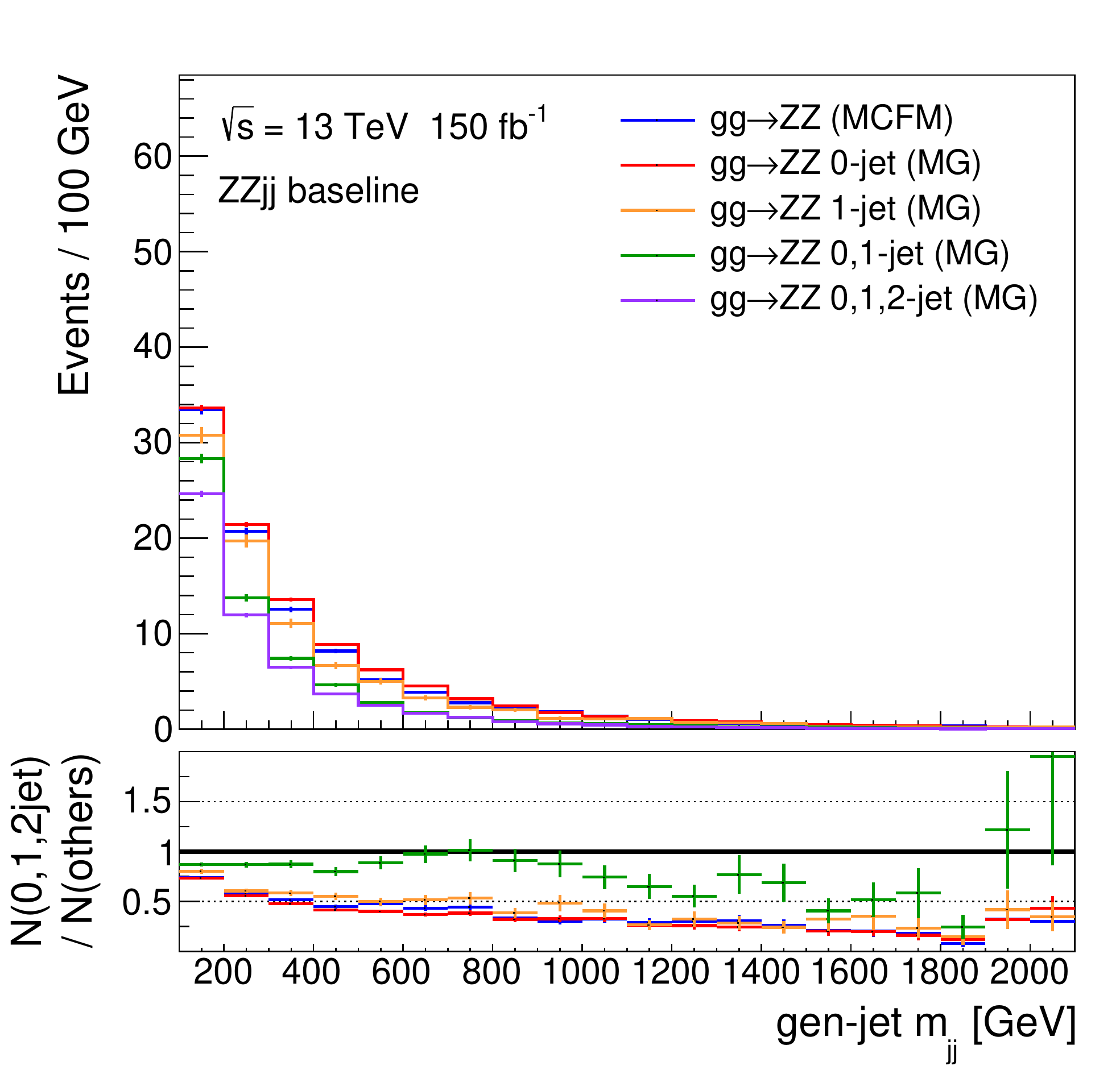}
\includegraphics[width=0.38\textwidth]{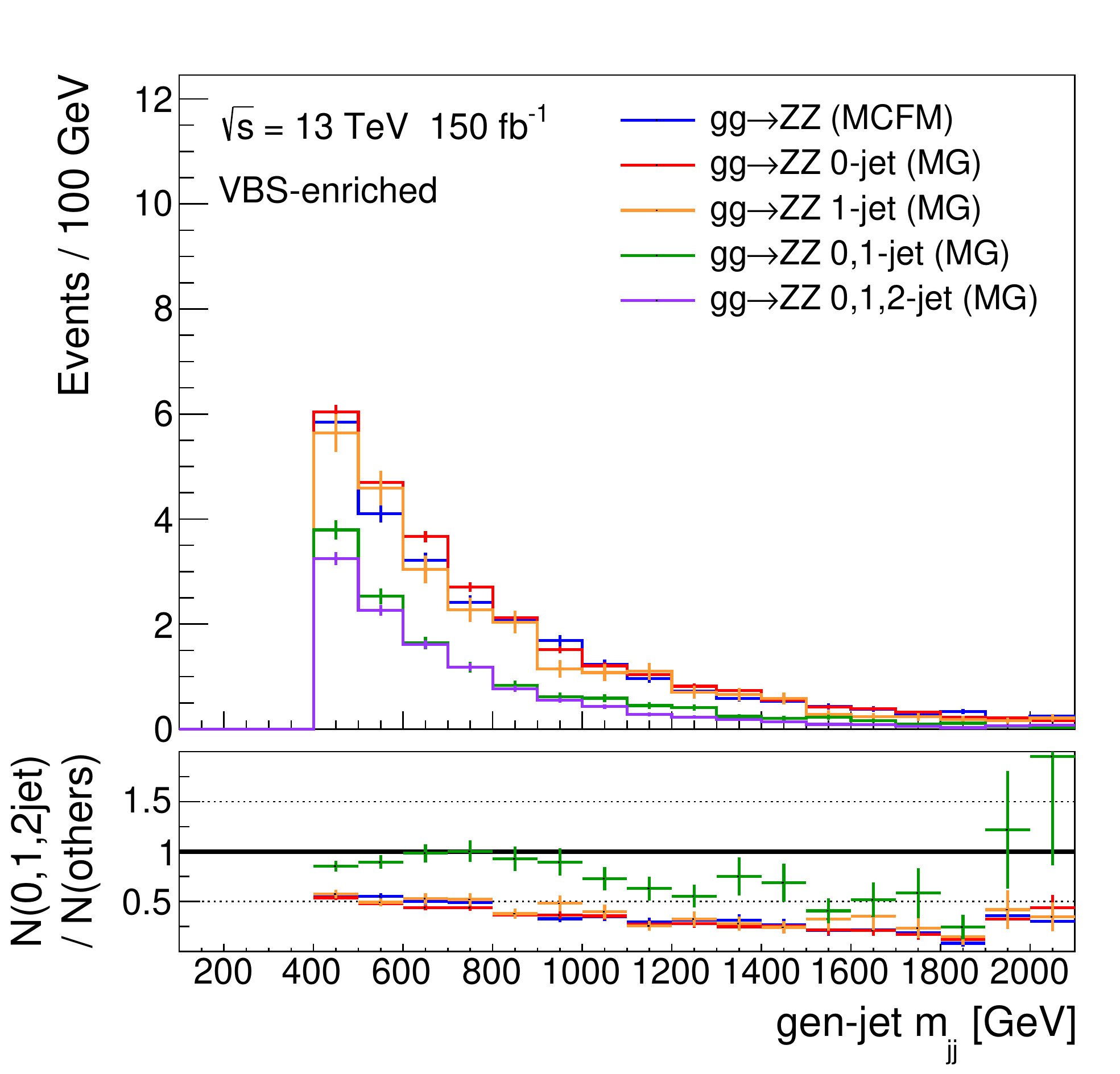}\\
\includegraphics[width=0.38\textwidth]{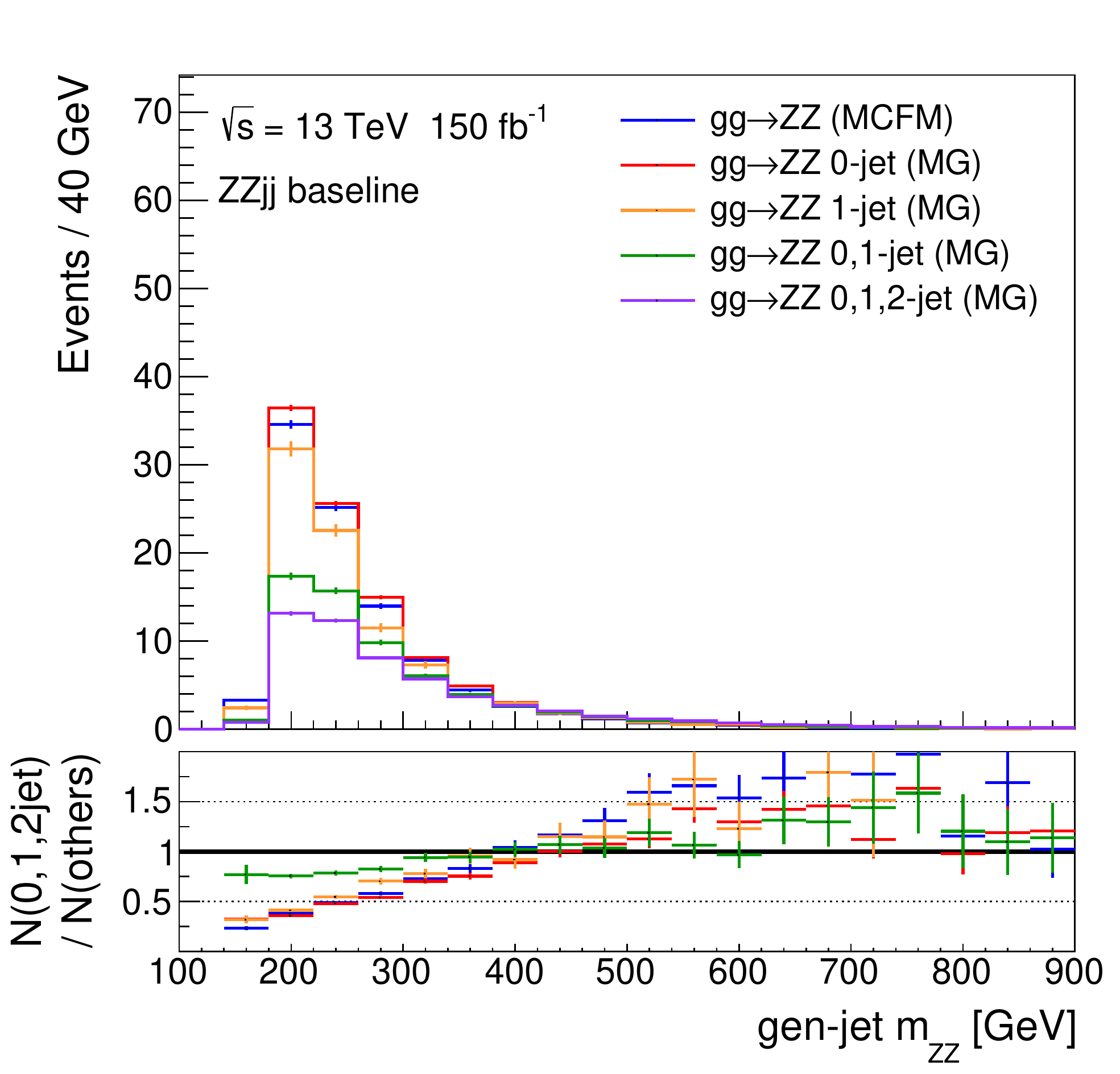}
\includegraphics[width=0.38\textwidth]{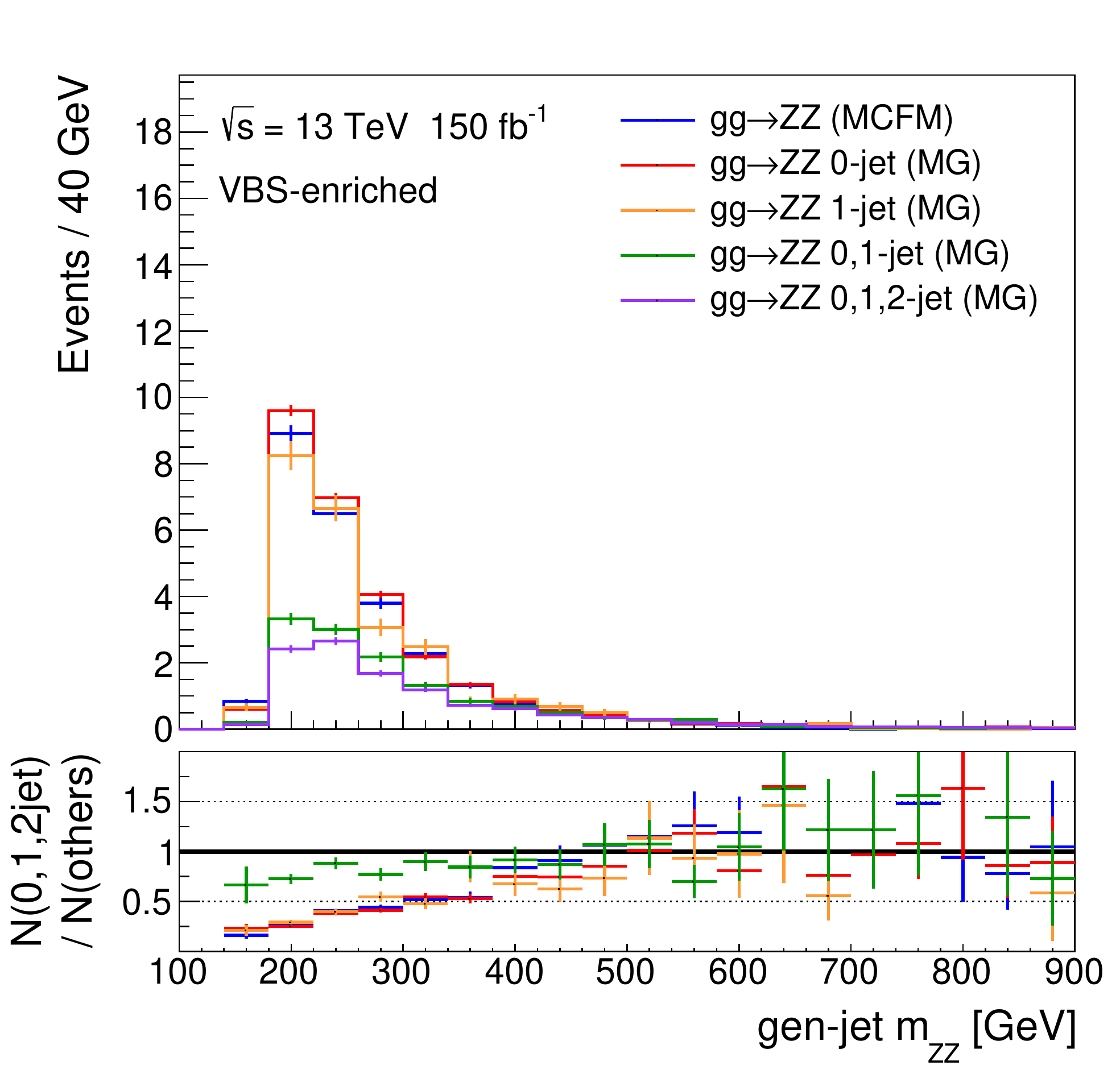}
\end{center}
\caption{Comparisons of generator-level kinematics $|\Delta\eta_{jj}|$, $m_{jj}$, and $m_{ZZ}$, among the \textsc{mcfm} $gg\to ZZ$ simulation (blue), the \textsc{MadGraph} $gg\to ZZ$ 0-jet simulation (red), 1-jet simulation (orange), $0,1$-jet MLM matched simulation (green), and the $0,1,2$-jet matched simulation (purple), passing the $ZZjj$ baseline selection (left) and the VBS-enriched selection (right) respectively. The error bars indicate the statistical uncertainties with the assumed integrated luminosity. All samples are normalized to the cross section obtained from \textsc{mcfm} after imposing the on-shell requirement, and correspond to an integrated luminosity of $150~\mathrm{fb}^{-1}$.}
\label{fig:reco_compare}
\end{figure*}

Fig.~\ref{fig:reco_compare} further shows differences in the distributions of the absolute dijet pseudorapidity separation $|\Delta \eta_{jj}|$, dijet invariant mass $m_{jj}$, and the mass of the $ZZ$ pair $m_{ZZ}$ among the four simulations, after the $ZZjj$ baseline and VBS-enriched selections. As can be seen, the new matched simulation gives lower event yields after the selections, consistent with the result in Table~\ref{tab:yields}. We summarize as follows:
\begin{itemize}
    \item As a consequence of validation results from Sec.~\ref{sec:val}, the softness of jets modeled in the MLM matched simulation may cause lower baseline selection efficiency, since a typical VBS region favors high-$p_{\mathrm{T}}$ jets. Hence the yields are generally smaller in each distribution of Fig.~\ref{fig:reco_compare} and in Table~\ref{tab:yields}. The compared 1-jet and $0,1$-jet matched simulations also show a decrease in event yields, but not as significant as for the $0,1,2$-jet case.
    \item As shown in the $|\Delta\eta_{jj}|$ distribution, the $0,1,2$-jet simulation with dijet simulated from matrix elements induces larger separation of jets, compared to the $0,1$-jet simulation where one jet is produced from matrix-element and another from parton showers, however the discrepancy in $|\Delta\eta_{jj}|$ vanishes after the VBS-enriched selection.
    \item Since the $ZZ$ pair recoils against the emitted jets, the softness of jets may in turn cause a larger $m_{ZZ}$. This explains the increasing ratio of yields in higher bins of the $m_{ZZ}$ distribution.
\end{itemize}

It is evident that the improvement in $gg\to ZZ$ simulation may impact the total background estimation, and thereafter influence the fit results of VBS signal searches.  We note that a similar behaviour may appear in other analyses with the change of jet description using improved loop-induced simulation.

\section{Conclusions}
In this study, we present for the first time the results of a fully exclusive simulation based on the matrix elements for loop-induced $ZZ + 0,1,2$ parton processes at leading order (LO) matched to parton showers. We find the optimal MLM matching cutoff scale to be smaller in this simulation compared to commonly used values. 
We examine and validate this new description by comparing it with various loop-induced $ZZ$ simulations, including a \textsc{mcfm} and \textsc{MadGraph} $ZZ$ simulations with jets from parton showers, a \textsc{MadGraph} simulation with $ZZ+1$-jet simulation, and an analogous \textsc{MadGraph} matched simulation for $0,1$-partons. We find that the $0,1,2$-parton matched simulation provides the most state-of-the-art exclusive description of the final state jets, despite its high complexity in event generation. Jets modeled from the $0,1,2$-parton matched description are found to exhibit a generally softer transverse momentum spectrum compared to pure parton-shower jets.

We also briefly discuss the physics impact on vector boson scattering (VBS) $ZZ$ measurements at the LHC. By replacing the earlier loop-induced $gg\to ZZ$ simulation with the new $0,1,2$-jet matched simulation, we observe a decrease of 43\% in event yields for a typical VBS $ZZjj$ baseline region, and of 56\% for a tighter VBS-enriched region. We also observe significant discrepancies in the generator-level jet or reconstructed $Z$ boson kinematics among the different modeling approaches. 
We hence suggest the implementation of a  more accurate description of the emitted jets in this process.

\acknowledgements
This work is supported in part by MOST under grant No.~2018YFA0403900, COST Action CA16108, and the Italy-Serbia MAE bilateral project No. RS19MO06. We thank the CNRS/IN2P3 and the France-China Particle Physics Laboratory (FCPPL) for their support.

\bibliographystyle{apsrev4-1}
\bibliography{loopzz}

\begin{thebibliography}{27}%
\makeatletter
\providecommand \@ifxundefined [1]{%
 \@ifx{#1\undefined}
}%
\providecommand \@ifnum [1]{%
 \ifnum #1\expandafter \@firstoftwo
 \else \expandafter \@secondoftwo
 \fi
}%
\providecommand \@ifx [1]{%
 \ifx #1\expandafter \@firstoftwo
 \else \expandafter \@secondoftwo
 \fi
}%
\providecommand \natexlab [1]{#1}%
\providecommand \enquote  [1]{``#1''}%
\providecommand \bibnamefont  [1]{#1}%
\providecommand \bibfnamefont [1]{#1}%
\providecommand \citenamefont [1]{#1}%
\providecommand \href@noop [0]{\@secondoftwo}%
\providecommand \href [0]{\begingroup \@sanitize@url \@href}%
\providecommand \@href[1]{\@@startlink{#1}\@@href}%
\providecommand \@@href[1]{\endgroup#1\@@endlink}%
\providecommand \@sanitize@url [0]{\catcode `\\12\catcode `\$12\catcode
  `\&12\catcode `\#12\catcode `\^12\catcode `\_12\catcode `\%12\relax}%
\providecommand \@@startlink[1]{}%
\providecommand \@@endlink[0]{}%
\providecommand \url  [0]{\begingroup\@sanitize@url \@url }%
\providecommand \@url [1]{\endgroup\@href {#1}{\urlprefix }}%
\providecommand \urlprefix  [0]{URL }%
\providecommand \Eprint [0]{\href }%
\providecommand \doibase [0]{http://dx.doi.org/}%
\providecommand \selectlanguage [0]{\@gobble}%
\providecommand \bibinfo  [0]{\@secondoftwo}%
\providecommand \bibfield  [0]{\@secondoftwo}%
\providecommand \translation [1]{[#1]}%
\providecommand \BibitemOpen [0]{}%
\providecommand \bibitemStop [0]{}%
\providecommand \bibitemNoStop [0]{.\EOS\space}%
\providecommand \EOS [0]{\spacefactor3000\relax}%
\providecommand \BibitemShut  [1]{\csname bibitem#1\endcsname}%
\let\auto@bib@innerbib\@empty
\bibitem [{\citenamefont {Binoth}\ \emph {et~al.}(2008)\citenamefont {Binoth},
  \citenamefont {Kauer},\ and\ \citenamefont {Mertsch}}]{Binoth:2008pr}%
  \BibitemOpen
  \bibfield  {author} {\bibinfo {author} {\bibfnamefont {T.}~\bibnamefont
  {Binoth}}, \bibinfo {author} {\bibfnamefont {N.}~\bibnamefont {Kauer}}, \
  and\ \bibinfo {author} {\bibfnamefont {P.}~\bibnamefont {Mertsch}},\ }in\
  \href {\doibase 10.3360/dis.2008.142} {\emph {\bibinfo {booktitle} {{16th
  International Workshop on Deep Inelastic Scattering and Related Subjects}}}}\
  (\bibinfo {year} {2008})\ p.\ \bibinfo {pages} {142},\ \Eprint
  {http://arxiv.org/abs/0807.0024} {arXiv:0807.0024 [hep-ph]} \BibitemShut
  {NoStop}%
\bibitem [{\citenamefont {Sirunyan}\ \emph
  {et~al.}(2017{\natexlab{a}})\citenamefont {Sirunyan} \emph
  {et~al.}}]{Sirunyan:2017exp}%
  \BibitemOpen
  \bibfield  {author} {\bibinfo {author} {\bibfnamefont {A.~M.}\ \bibnamefont
  {Sirunyan}} \emph {et~al.} (\bibinfo {collaboration} {CMS}),\ }\href
  {\doibase 10.1007/JHEP11(2017)047} {\bibfield  {journal} {\bibinfo  {journal}
  {JHEP}\ }\textbf {\bibinfo {volume} {11}},\ \bibinfo {pages} {047} (\bibinfo
  {year} {2017}{\natexlab{a}})},\ \Eprint {http://arxiv.org/abs/1706.09936}
  {arXiv:1706.09936 [hep-ex]} \BibitemShut {NoStop}%
\bibitem [{\citenamefont {Aaboud}\ \emph {et~al.}(2017)\citenamefont {Aaboud}
  \emph {et~al.}}]{Aaboud:2017oem}%
  \BibitemOpen
  \bibfield  {author} {\bibinfo {author} {\bibfnamefont {M.}~\bibnamefont
  {Aaboud}} \emph {et~al.} (\bibinfo {collaboration} {ATLAS}),\ }\href
  {\doibase 10.1007/JHEP10(2017)132} {\bibfield  {journal} {\bibinfo  {journal}
  {JHEP}\ }\textbf {\bibinfo {volume} {10}},\ \bibinfo {pages} {132} (\bibinfo
  {year} {2017})},\ \Eprint {http://arxiv.org/abs/1708.02810} {arXiv:1708.02810
  [hep-ex]} \BibitemShut {NoStop}%
\bibitem [{\citenamefont {Aaboud}\ \emph
  {et~al.}(2018{\natexlab{a}})\citenamefont {Aaboud} \emph
  {et~al.}}]{Aaboud:2018puo}%
  \BibitemOpen
  \bibfield  {author} {\bibinfo {author} {\bibfnamefont {M.}~\bibnamefont
  {Aaboud}} \emph {et~al.} (\bibinfo {collaboration} {ATLAS}),\ }\href
  {\doibase 10.1016/j.physletb.2018.09.048} {\bibfield  {journal} {\bibinfo
  {journal} {Phys. Lett. B}\ }\textbf {\bibinfo {volume} {786}},\ \bibinfo
  {pages} {223} (\bibinfo {year} {2018}{\natexlab{a}})},\ \Eprint
  {http://arxiv.org/abs/1808.01191} {arXiv:1808.01191 [hep-ex]} \BibitemShut
  {NoStop}%
\bibitem [{\citenamefont {Sirunyan}\ \emph
  {et~al.}(2019{\natexlab{a}})\citenamefont {Sirunyan} \emph
  {et~al.}}]{Sirunyan:2019twz}%
  \BibitemOpen
  \bibfield  {author} {\bibinfo {author} {\bibfnamefont {A.~M.}\ \bibnamefont
  {Sirunyan}} \emph {et~al.} (\bibinfo {collaboration} {CMS}),\ }\href
  {\doibase 10.1103/PhysRevD.99.112003} {\bibfield  {journal} {\bibinfo
  {journal} {Phys. Rev. D}\ }\textbf {\bibinfo {volume} {99}},\ \bibinfo
  {pages} {112003} (\bibinfo {year} {2019}{\natexlab{a}})},\ \Eprint
  {http://arxiv.org/abs/1901.00174} {arXiv:1901.00174 [hep-ex]} \BibitemShut
  {NoStop}%
\bibitem [{\citenamefont {Aaboud}\ \emph
  {et~al.}(2018{\natexlab{b}})\citenamefont {Aaboud} \emph
  {et~al.}}]{Aaboud:2017rwm}%
  \BibitemOpen
  \bibfield  {author} {\bibinfo {author} {\bibfnamefont {M.}~\bibnamefont
  {Aaboud}} \emph {et~al.} (\bibinfo {collaboration} {ATLAS}),\ }\href
  {\doibase 10.1103/PhysRevD.97.032005} {\bibfield  {journal} {\bibinfo
  {journal} {Phys. Rev. D}\ }\textbf {\bibinfo {volume} {97}},\ \bibinfo
  {pages} {032005} (\bibinfo {year} {2018}{\natexlab{b}})},\ \Eprint
  {http://arxiv.org/abs/1709.07703} {arXiv:1709.07703 [hep-ex]} \BibitemShut
  {NoStop}%
\bibitem [{\citenamefont {Sirunyan}\ \emph
  {et~al.}(2019{\natexlab{b}})\citenamefont {Sirunyan} \emph
  {et~al.}}]{Sirunyan:2018vkx}%
  \BibitemOpen
  \bibfield  {author} {\bibinfo {author} {\bibfnamefont {A.~M.}\ \bibnamefont
  {Sirunyan}} \emph {et~al.} (\bibinfo {collaboration} {CMS}),\ }\href
  {\doibase 10.1016/j.physletb.2018.11.007} {\bibfield  {journal} {\bibinfo
  {journal} {Phys. Lett. B}\ }\textbf {\bibinfo {volume} {789}},\ \bibinfo
  {pages} {19} (\bibinfo {year} {2019}{\natexlab{b}})},\ \Eprint
  {http://arxiv.org/abs/1806.11073} {arXiv:1806.11073 [hep-ex]} \BibitemShut
  {NoStop}%
\bibitem [{\citenamefont {Sirunyan}\ \emph {et~al.}(2020)\citenamefont
  {Sirunyan} \emph {et~al.}}]{Sirunyan:2020alo}%
  \BibitemOpen
  \bibfield  {author} {\bibinfo {author} {\bibfnamefont {A.~M.}\ \bibnamefont
  {Sirunyan}} \emph {et~al.} (\bibinfo {collaboration} {CMS}),\ }\href@noop {}
  {\  (\bibinfo {year} {2020})},\ \Eprint {http://arxiv.org/abs/2008.07013}
  {arXiv:2008.07013 [hep-ex]} \BibitemShut {NoStop}%
\bibitem [{\citenamefont {Aad}\ \emph {et~al.}(2020)\citenamefont {Aad} \emph
  {et~al.}}]{Aad:2020zbq}%
  \BibitemOpen
  \bibfield  {author} {\bibinfo {author} {\bibfnamefont {G.}~\bibnamefont
  {Aad}} \emph {et~al.} (\bibinfo {collaboration} {ATLAS}),\ }\href@noop {} {\
  (\bibinfo {year} {2020})},\ \Eprint {http://arxiv.org/abs/2004.10612}
  {arXiv:2004.10612 [hep-ex]} \BibitemShut {NoStop}%
\bibitem [{\citenamefont {Khachatryan}\ \emph {et~al.}(2017)\citenamefont
  {Khachatryan} \emph {et~al.}}]{Khachatryan:2017wny}%
  \BibitemOpen
  \bibfield  {author} {\bibinfo {author} {\bibfnamefont {V.}~\bibnamefont
  {Khachatryan}} \emph {et~al.} (\bibinfo {collaboration} {CMS}),\ }\href
  {\doibase 10.1016/j.physletb.2017.08.069} {\bibfield  {journal} {\bibinfo
  {journal} {Phys. Lett. B}\ }\textbf {\bibinfo {volume} {773}},\ \bibinfo
  {pages} {563} (\bibinfo {year} {2017})},\ \Eprint
  {http://arxiv.org/abs/1701.01345} {arXiv:1701.01345 [hep-ex]} \BibitemShut
  {NoStop}%
\bibitem [{\citenamefont {Aaboud}\ \emph
  {et~al.}(2018{\natexlab{c}})\citenamefont {Aaboud} \emph
  {et~al.}}]{Aaboud:2017rel}%
  \BibitemOpen
  \bibfield  {author} {\bibinfo {author} {\bibfnamefont {M.}~\bibnamefont
  {Aaboud}} \emph {et~al.} (\bibinfo {collaboration} {ATLAS}),\ }\href
  {\doibase 10.1140/epjc/s10052-018-5686-3} {\bibfield  {journal} {\bibinfo
  {journal} {Eur. Phys. J. C}\ }\textbf {\bibinfo {volume} {78}},\ \bibinfo
  {pages} {293} (\bibinfo {year} {2018}{\natexlab{c}})},\ \Eprint
  {http://arxiv.org/abs/1712.06386} {arXiv:1712.06386 [hep-ex]} \BibitemShut
  {NoStop}%
\bibitem [{\citenamefont {Caola}\ \emph {et~al.}(2015)\citenamefont {Caola},
  \citenamefont {Melnikov}, \citenamefont {{R\"ontsch}},\ and\ \citenamefont
  {Tancredi}}]{Caola:2015psa}%
  \BibitemOpen
  \bibfield  {author} {\bibinfo {author} {\bibfnamefont {F.}~\bibnamefont
  {Caola}}, \bibinfo {author} {\bibfnamefont {K.}~\bibnamefont {Melnikov}},
  \bibinfo {author} {\bibfnamefont {R.}~\bibnamefont {{R\"ontsch}}}, \ and\
  \bibinfo {author} {\bibfnamefont {L.}~\bibnamefont {Tancredi}},\ }\href
  {\doibase 10.1103/PhysRevD.92.094028} {\bibfield  {journal} {\bibinfo
  {journal} {Phys. Rev. D}\ }\textbf {\bibinfo {volume} {92}},\ \bibinfo
  {pages} {094028} (\bibinfo {year} {2015})},\ \Eprint
  {http://arxiv.org/abs/1509.06734} {arXiv:1509.06734 [hep-ph]} \BibitemShut
  {NoStop}%
\bibitem [{\citenamefont {Alioli}\ \emph {et~al.}(2017)\citenamefont {Alioli},
  \citenamefont {Caola}, \citenamefont {Luisoni},\ and\ \citenamefont
  {{R\"ontsch}}}]{Alioli:2016xab}%
  \BibitemOpen
  \bibfield  {author} {\bibinfo {author} {\bibfnamefont {S.}~\bibnamefont
  {Alioli}}, \bibinfo {author} {\bibfnamefont {F.}~\bibnamefont {Caola}},
  \bibinfo {author} {\bibfnamefont {G.}~\bibnamefont {Luisoni}}, \ and\
  \bibinfo {author} {\bibfnamefont {R.}~\bibnamefont {{R\"ontsch}}},\ }\href
  {\doibase 10.1103/PhysRevD.95.034042} {\bibfield  {journal} {\bibinfo
  {journal} {Phys. Rev. D}\ }\textbf {\bibinfo {volume} {95}},\ \bibinfo
  {pages} {034042} (\bibinfo {year} {2017})},\ \Eprint
  {http://arxiv.org/abs/1609.09719} {arXiv:1609.09719 [hep-ph]} \BibitemShut
  {NoStop}%
\bibitem [{\citenamefont {Caola}\ \emph {et~al.}(2016)\citenamefont {Caola},
  \citenamefont {Dowling}, \citenamefont {Melnikov}, \citenamefont
  {{R\"ontsch}},\ and\ \citenamefont {Tancredi}}]{Caola:2016trd}%
  \BibitemOpen
  \bibfield  {author} {\bibinfo {author} {\bibfnamefont {F.}~\bibnamefont
  {Caola}}, \bibinfo {author} {\bibfnamefont {M.}~\bibnamefont {Dowling}},
  \bibinfo {author} {\bibfnamefont {K.}~\bibnamefont {Melnikov}}, \bibinfo
  {author} {\bibfnamefont {R.}~\bibnamefont {{R\"ontsch}}}, \ and\ \bibinfo
  {author} {\bibfnamefont {L.}~\bibnamefont {Tancredi}},\ }\href {\doibase
  10.1007/JHEP07(2016)087} {\bibfield  {journal} {\bibinfo  {journal} {JHEP}\
  }\textbf {\bibinfo {volume} {07}},\ \bibinfo {pages} {087} (\bibinfo {year}
  {2016})},\ \Eprint {http://arxiv.org/abs/1605.04610} {arXiv:1605.04610
  [hep-ph]} \BibitemShut {NoStop}%
\bibitem [{\citenamefont {Grazzini}\ \emph {et~al.}(2019)\citenamefont
  {Grazzini}, \citenamefont {Kallweit}, \citenamefont {Wiesemann},\ and\
  \citenamefont {Yook}}]{Grazzini:2018owa}%
  \BibitemOpen
  \bibfield  {author} {\bibinfo {author} {\bibfnamefont {M.}~\bibnamefont
  {Grazzini}}, \bibinfo {author} {\bibfnamefont {S.}~\bibnamefont {Kallweit}},
  \bibinfo {author} {\bibfnamefont {M.}~\bibnamefont {Wiesemann}}, \ and\
  \bibinfo {author} {\bibfnamefont {J.~Y.}\ \bibnamefont {Yook}},\ }\href
  {\doibase 10.1007/JHEP03(2019)070} {\bibfield  {journal} {\bibinfo  {journal}
  {JHEP}\ }\textbf {\bibinfo {volume} {03}},\ \bibinfo {pages} {070} (\bibinfo
  {year} {2019})},\ \Eprint {http://arxiv.org/abs/1811.09593} {arXiv:1811.09593
  [hep-ph]} \BibitemShut {NoStop}%
\bibitem [{\citenamefont {Campanario}\ \emph {et~al.}(2013)\citenamefont
  {Campanario}, \citenamefont {Li}, \citenamefont {Rauch},\ and\ \citenamefont
  {Spira}}]{Campanario:2012bh}%
  \BibitemOpen
  \bibfield  {author} {\bibinfo {author} {\bibfnamefont {F.}~\bibnamefont
  {Campanario}}, \bibinfo {author} {\bibfnamefont {Q.}~\bibnamefont {Li}},
  \bibinfo {author} {\bibfnamefont {M.}~\bibnamefont {Rauch}}, \ and\ \bibinfo
  {author} {\bibfnamefont {M.}~\bibnamefont {Spira}},\ }\href {\doibase
  10.1007/JHEP06(2013)069} {\bibfield  {journal} {\bibinfo  {journal} {JHEP}\
  }\textbf {\bibinfo {volume} {06}},\ \bibinfo {pages} {069} (\bibinfo {year}
  {2013})},\ \Eprint {http://arxiv.org/abs/1211.5429} {arXiv:1211.5429
  [hep-ph]} \BibitemShut {NoStop}%
\bibitem [{\citenamefont {Catani}\ \emph {et~al.}(2001)\citenamefont {Catani},
  \citenamefont {Krauss}, \citenamefont {Kuhn},\ and\ \citenamefont
  {Webber}}]{Catani:2001cc}%
  \BibitemOpen
  \bibfield  {author} {\bibinfo {author} {\bibfnamefont {S.}~\bibnamefont
  {Catani}}, \bibinfo {author} {\bibfnamefont {F.}~\bibnamefont {Krauss}},
  \bibinfo {author} {\bibfnamefont {R.}~\bibnamefont {Kuhn}}, \ and\ \bibinfo
  {author} {\bibfnamefont {B.}~\bibnamefont {Webber}},\ }\href {\doibase
  10.1088/1126-6708/2001/11/063} {\bibfield  {journal} {\bibinfo  {journal}
  {JHEP}\ }\textbf {\bibinfo {volume} {11}},\ \bibinfo {pages} {063} (\bibinfo
  {year} {2001})},\ \Eprint {http://arxiv.org/abs/hep-ph/0109231}
  {arXiv:hep-ph/0109231} \BibitemShut {NoStop}%
\bibitem [{\citenamefont {Alwall}\ \emph {et~al.}(2008)\citenamefont {Alwall}
  \emph {et~al.}}]{Alwall:2007fs}%
  \BibitemOpen
  \bibfield  {author} {\bibinfo {author} {\bibfnamefont {J.}~\bibnamefont
  {Alwall}} \emph {et~al.},\ }\href {\doibase 10.1140/epjc/s10052-007-0490-5}
  {\bibfield  {journal} {\bibinfo  {journal} {Eur. Phys. J. C}\ }\textbf
  {\bibinfo {volume} {53}},\ \bibinfo {pages} {473} (\bibinfo {year} {2008})},\
  \Eprint {http://arxiv.org/abs/0706.2569} {arXiv:0706.2569 [hep-ph]}
  \BibitemShut {NoStop}%
\bibitem [{\citenamefont {Cascioli}\ \emph {et~al.}(2014)\citenamefont
  {Cascioli}, \citenamefont {H\"oche}, \citenamefont {Krauss}, \citenamefont
  {Maierh\"ofer}, \citenamefont {Pozzorini},\ and\ \citenamefont
  {Siegert}}]{Cascioli:2013gfa}%
  \BibitemOpen
  \bibfield  {author} {\bibinfo {author} {\bibfnamefont {F.}~\bibnamefont
  {Cascioli}}, \bibinfo {author} {\bibfnamefont {S.}~\bibnamefont {H\"oche}},
  \bibinfo {author} {\bibfnamefont {F.}~\bibnamefont {Krauss}}, \bibinfo
  {author} {\bibfnamefont {P.}~\bibnamefont {Maierh\"ofer}}, \bibinfo {author}
  {\bibfnamefont {S.}~\bibnamefont {Pozzorini}}, \ and\ \bibinfo {author}
  {\bibfnamefont {F.}~\bibnamefont {Siegert}},\ }\href {\doibase
  10.1007/JHEP01(2014)046} {\bibfield  {journal} {\bibinfo  {journal} {JHEP}\
  }\textbf {\bibinfo {volume} {01}},\ \bibinfo {pages} {046} (\bibinfo {year}
  {2014})},\ \Eprint {http://arxiv.org/abs/1309.0500} {arXiv:1309.0500
  [hep-ph]} \BibitemShut {NoStop}%
\bibitem [{\citenamefont {Alwall}\ \emph {et~al.}(2014)\citenamefont {Alwall},
  \citenamefont {Frederix}, \citenamefont {Frixione}, \citenamefont {Hirschi},
  \citenamefont {Maltoni}, \citenamefont {Mattelaer}, \citenamefont {Shao},
  \citenamefont {Stelzer}, \citenamefont {Torrielli},\ and\ \citenamefont
  {Zaro}}]{Alwall:2014hca}%
  \BibitemOpen
  \bibfield  {author} {\bibinfo {author} {\bibfnamefont {J.}~\bibnamefont
  {Alwall}}, \bibinfo {author} {\bibfnamefont {R.}~\bibnamefont {Frederix}},
  \bibinfo {author} {\bibfnamefont {S.}~\bibnamefont {Frixione}}, \bibinfo
  {author} {\bibfnamefont {V.}~\bibnamefont {Hirschi}}, \bibinfo {author}
  {\bibfnamefont {F.}~\bibnamefont {Maltoni}}, \bibinfo {author} {\bibfnamefont
  {O.}~\bibnamefont {Mattelaer}}, \bibinfo {author} {\bibfnamefont {H.~S.}\
  \bibnamefont {Shao}}, \bibinfo {author} {\bibfnamefont {T.}~\bibnamefont
  {Stelzer}}, \bibinfo {author} {\bibfnamefont {P.}~\bibnamefont {Torrielli}},
  \ and\ \bibinfo {author} {\bibfnamefont {M.}~\bibnamefont {Zaro}},\ }\href
  {\doibase 10.1007/JHEP07(2014)079} {\bibfield  {journal} {\bibinfo  {journal}
  {JHEP}\ }\textbf {\bibinfo {volume} {07}},\ \bibinfo {pages} {079} (\bibinfo
  {year} {2014})},\ \Eprint {http://arxiv.org/abs/1405.0301} {arXiv:1405.0301
  [hep-ph]} \BibitemShut {NoStop}%
\bibitem [{\citenamefont {Hirschi}\ and\ \citenamefont
  {Mattelaer}(2015)}]{Hirschi:2015iia}%
  \BibitemOpen
  \bibfield  {author} {\bibinfo {author} {\bibfnamefont {V.}~\bibnamefont
  {Hirschi}}\ and\ \bibinfo {author} {\bibfnamefont {O.}~\bibnamefont
  {Mattelaer}},\ }\href {\doibase 10.1007/JHEP10(2015)146} {\bibfield
  {journal} {\bibinfo  {journal} {JHEP}\ }\textbf {\bibinfo {volume} {10}},\
  \bibinfo {pages} {146} (\bibinfo {year} {2015})},\ \Eprint
  {http://arxiv.org/abs/1507.00020} {arXiv:1507.00020 [hep-ph]} \BibitemShut
  {NoStop}%
\bibitem [{MLM(2016)}]{MLMDISCUSS}%
  \BibitemOpen
  \href@noop {} {\enquote {\bibinfo {title} {{\textsc{MadGraph5\_aMC@NLO}
  launchpad}},}\ }\bibinfo {howpublished}
  {\url{https://answers.launchpad.net/mg5amcnlo/+question/402723}} (\bibinfo
  {year} {2016})\BibitemShut {NoStop}%
\bibitem [{\citenamefont {Alwall}\ \emph {et~al.}(2009)\citenamefont {Alwall},
  \citenamefont {de~Visscher},\ and\ \citenamefont {Maltoni}}]{Alwall:2008qv}%
  \BibitemOpen
  \bibfield  {author} {\bibinfo {author} {\bibfnamefont {J.}~\bibnamefont
  {Alwall}}, \bibinfo {author} {\bibfnamefont {S.}~\bibnamefont {de~Visscher}},
  \ and\ \bibinfo {author} {\bibfnamefont {F.}~\bibnamefont {Maltoni}},\ }\href
  {\doibase 10.1088/1126-6708/2009/02/017} {\bibfield  {journal} {\bibinfo
  {journal} {JHEP}\ }\textbf {\bibinfo {volume} {02}},\ \bibinfo {pages} {017}
  (\bibinfo {year} {2009})},\ \Eprint {http://arxiv.org/abs/0810.5350}
  {arXiv:0810.5350 [hep-ph]} \BibitemShut {NoStop}%
\bibitem [{MAT(2013)}]{MATCHINGSCALE}%
  \BibitemOpen
  \href@noop {} {\enquote {\bibinfo {title} {{\textsc{MadGraph5\_aMC@NLO}
  wiki}},}\ }\bibinfo {howpublished}
  {\url{https://cp3.irmp.ucl.ac.be/projects/madgraph/wiki/IntroMatching}}
  (\bibinfo {year} {2013})\BibitemShut {NoStop}%
\bibitem [{SCA(2012)}]{SCALECHOICE}%
  \BibitemOpen
  \href@noop {} {\enquote {\bibinfo {title} {{\textsc{MadGraph5\_aMC@NLO}
  wiki}},}\ }\bibinfo {howpublished}
  {\url{https://cp3.irmp.ucl.ac.be/projects/madgraph/wiki/FAQ-General-13}}
  (\bibinfo {year} {2012})\BibitemShut {NoStop}%
\bibitem [{\citenamefont {Boughezal}\ \emph {et~al.}(2017)\citenamefont
  {Boughezal}, \citenamefont {Campbell}, \citenamefont {Ellis}, \citenamefont
  {Focke}, \citenamefont {Giele}, \citenamefont {Liu}, \citenamefont
  {Petriello},\ and\ \citenamefont {Williams}}]{Boughezal:2016wmq}%
  \BibitemOpen
  \bibfield  {author} {\bibinfo {author} {\bibfnamefont {R.}~\bibnamefont
  {Boughezal}}, \bibinfo {author} {\bibfnamefont {J.~M.}\ \bibnamefont
  {Campbell}}, \bibinfo {author} {\bibfnamefont {R.~K.}\ \bibnamefont {Ellis}},
  \bibinfo {author} {\bibfnamefont {C.}~\bibnamefont {Focke}}, \bibinfo
  {author} {\bibfnamefont {W.}~\bibnamefont {Giele}}, \bibinfo {author}
  {\bibfnamefont {X.}~\bibnamefont {Liu}}, \bibinfo {author} {\bibfnamefont
  {F.}~\bibnamefont {Petriello}}, \ and\ \bibinfo {author} {\bibfnamefont
  {C.}~\bibnamefont {Williams}},\ }\href {\doibase
  10.1140/epjc/s10052-016-4558-y} {\bibfield  {journal} {\bibinfo  {journal}
  {Eur. Phys. J. C}\ }\textbf {\bibinfo {volume} {77}},\ \bibinfo {pages} {7}
  (\bibinfo {year} {2017})},\ \Eprint {http://arxiv.org/abs/1605.08011}
  {arXiv:1605.08011 [hep-ph]} \BibitemShut {NoStop}%
\bibitem [{\citenamefont {Sirunyan}\ \emph
  {et~al.}(2017{\natexlab{b}})\citenamefont {Sirunyan} \emph
  {et~al.}}]{Sirunyan:2017fvv}%
  \BibitemOpen
  \bibfield  {author} {\bibinfo {author} {\bibfnamefont {A.~M.}\ \bibnamefont
  {Sirunyan}} \emph {et~al.} (\bibinfo {collaboration} {CMS}),\ }\href
  {\doibase 10.1016/j.physletb.2017.10.020} {\bibfield  {journal} {\bibinfo
  {journal} {Phys. Lett. B}\ }\textbf {\bibinfo {volume} {774}},\ \bibinfo
  {pages} {682} (\bibinfo {year} {2017}{\natexlab{b}})},\ \Eprint
  {http://arxiv.org/abs/1708.02812} {arXiv:1708.02812 [hep-ex]} \BibitemShut
  {NoStop}%
\end{thebibliography}%
\end{document}